\documentclass{aa}  

\usepackage{graphicx}
\usepackage{txfonts}
\usepackage[colorlinks,allcolors=blue,breaklinks=true]{hyperref}
\usepackage{longtable}
\usepackage{rotating}
\usepackage{lscape}
\usepackage{soul}
\usepackage{xcolor}
\usepackage{multirow}
\usepackage{graphicx}
\usepackage{tcolorbox}
\usepackage{enumitem}

\begin{document}

   \title{The NEMESIS Catalogue of Young Stellar Objects for the Orion Star Formation Complex}
     \subtitle{I. General description of data curation\thanks{Our database can be accessed at \url{https://drive.google.com/drive/folders/1h6_dNt3It6T6xMyCyc5ssKlf9YDIkW8T?}}}

\author{J. Roquette
          \inst{1}
          \and
           M. Audard
          \inst{1}     
          \and
          D. Hernandez\inst{2}          
          \and    
          I. Gezer \inst{3,4}
          \and
          G. Marton\inst{3,4}
          \and
          C. Mas\inst{1}
          \and
          M. Madarász\inst{3,4}
          \and
          O. Dionatos \inst{5}
          }
    \institute{
   Department of Astronomy, University of Geneva, Chemin Pegasi 51,
1290 Versoix, Switzerland\\
              \email{julia.roquette@unige.ch}
               \and
            Universität Wien, Institut für Astrophysik, T\"urkenschanzstrasse 17, 1180 Wien, Austria
            \and
            Konkoly Observatory, Research Centre for Astronomy and Earth Sciences, Hungarian Research Network (HUN-REN), Konkoly Thege Mikl\'{o}s \'{U}t 15-17, 1121 Budapest, Hungary
            \and
            CSFK, MTA Centre of Excellence, Budapest, Konkoly Thege Miklós út 15-17., H-1121, Hungary
             \and
             Natural History Museum Vienna, Burgring 7, 1010 Vienna, Austria
}

   \date{Received August 29, 1997; accepted August 29, 1997}

   \abstract
   {The past decade has seen a rise in the use of machine learning methods in the study of young stellar evolution. This trend has led to a growing need for a comprehensive database of young stellar objects (YSO) that goes beyond survey-specific biases and that can be employed for training, validation, and refining the physical interpretation of machine learning outcomes.}
   {We aimed to review the literature focused on the Orion Star Formation complex (OSFC) to compile a thorough catalogue of previously identified YSO candidates in the region including the curation of observables relevant to probe their youth.}
   {Starting from the NASA/ADS database, we assembled YSO candidates from more than 200 peer-reviewed publications targeting the OSFC. We collated data products relevant to the study of young stars into a dedicated catalogue, which was complemented with data from large photometric and spectroscopic surveys and in the Strasbourg astronomical Data Center. We also added significant value to the catalogue by homogeneously deriving YSO infrared classification labels and through a comprehensive curation of labels concerning sources' multiplicity. Finally, we used a panchromatic approach to derive the probabilities that the candidate YSOs in our catalogue were contaminant extragalactic sources or giant stars.}
   {We present the NEMESIS catalogue of YSOs for the OSFC, which includes data collated for 27\,879 sources covering the whole mass spectrum and the various stages of pre-Main Sequence evolution from protostars to discless young stars. The catalogue includes a large collection of panchromatic photometric data processed into spectral energy distributions, stellar parameters ($T_\mathrm{eff}, T_\mathrm{bol}$, spectral types, $\log{g}$, $v\sin{i}$, RV), infrared classes, equivalent widths of emission lines related to YSOs accretion and star-disc interaction, and absorption lines such as lithium and lines related to source's gravity, X-ray emission observables, photometric variability observables (e.g., variability periods and amplitudes), and multiplicity labels.}
   {}


   \maketitle
   \nolinenumbers
%
\section{Introduction}

As astronomy dives into the era of Big Data, we are now faced with the challenges of reliable applications of artificial intelligence and machine learning techniques. A major obstacle for unbiased machine learning applications in the domain of star formation stems from the data available to validate results, which are often susceptible to small numbers or limited to the specifications of a few larger-scale surveys. To address the scarcity of polyvalent databases of young stars, we revisited the published literature targeting the Orion star-formation complex (OSFC) under the framework of the NEMESIS to build a reference sample of Young Stellar Objects (YSOs) for use in upcoming research. We hereby report the work for data curation and a general description of the NEMESIS catalogue of young stellar objects for the Orion Star Formation Complex.

NEMESIS (New Evolutionary Model for Early stages of Stars with Intelligent Systems) is an H2020 project\footnote{\url{https://nemesis.konkoly.hu/}} aiming at revisiting the star formation paradigm with the aid of Big Data and machine learning techniques. NEMESIS efforts included a new version of the Herschel/PACS point source catalogue \citep{Marton2024AA...688A.203M} with an associated Deep Neural Network approach for source removal \citep{NEMESIS_Mate}; the study of the 2D morphology of YSOs in photometric images using self-organising maps \citep{NEMESIS_David}, the variability characterisation of an all-sky sample YSOs from Gaia DR3 \citep{NEMESIS_Chloe}, a compendium of the application of deep learning techniques for the identification of YSOs all sky \citep{Marton2024eas..conf.1922M, NEMESIS_YSO_Gabor} with the three latter presenting applications of the present catalogue. Additionally, two follow-up studies added value to the present catalogue through the fitting of YSO's spectral energy distributions (SEDs) to synthetic libraries based on radiative transfer models \citep{NEMESIS_Ilknur}, and the ranking of bona fide YSOs (Roquette et al. in prep. 2025).

The Orion Star Formation Complex \citep[OSFC;][]{Bally2008hsf1.book..459B} was chosen as our target as it is the largest and most diverse nearby star formation region within $\lesssim$500 pc of the Sun, with tenths of thousands of YSOs distributed over almost 600 square degrees in the sky (see Fig.~\ref{fig:Orion_FOV}). The OSFC harbours populations that span all ages relevant to studying young stellar evolution \citep[$\lesssim 12$  Myr;][]{2018AJ....156...84K} and covers the whole stellar mass spectrum from high-mass stars to intermediate- and low-mass YSOs \citep[e.g.,][]{1997AJ....113.1733H,Muench2008hsf1.book..483M,ODell2008hsf1.book..544O}. 
 Previous research reported relatively large samples of YSOs located in the OSFC, but these were susceptible to the methodological specificities and limitations inherent to certain instruments or focused on subregions. For example, studies carried out with the \emph{Spitzer} space telescope \citep[e.g.,][]{2012AJ....144..192M,2016AJ....151....5M,2013ApJS..207....5F,2018MNRAS.477..298G} collectively identified $\sim10\,000$ YSOs. However, these are limited by \emph{Spitzer}'s reduced coverage of the OSFC and favour the detection of less evolved YSOs with significant contributions of a disc or envelope to their SEDs. Large-scale kinematic surveys based on \emph{Gaia} \citep[e.g.][]{2018AandA...620A.172Z} characterised a similar number of YSOs, but are conversely biased towards optically visible sources. Similarly, large-scale spectroscopic surveys \citep[e.g.,][]{2016ApJ...818...59D,2018AJ....156...84K,2019AJ....157...85B,2023AJ....165..205H} focussed on the brightest sources located in less crowded areas and are accordingly biased toward more evolved YSOs, which already emerged from their envelopes, being near-infrared bright and optically visible. With the NEMESIS Catalogue of YSOs for the OSFC, we propose to go beyond these limitations by compiling a cumulative source list from the ensemble of previous research and combining complementation from public datasets and modern data science techniques to and mitigate, as possible, the biases from previous surveys. 

Our data curation (Sect.~\ref{sec:data}) started from a historical data compilation based on peer-reviewed articles focused on studying YSOs in the OSFC (Sect.~\ref{sec:data-historical}). This historical compilation was then complemented with data from large photometric and spectroscopic surveys 
(Sect.~\ref{sec:largePhotSurveys} and \ref{sec:spec-surveys}) and from photometric data available at the Centre de données astronomiques de Strasbourg \citep[CDS;][Sect.~\ref{sec:VizierSED}]{Genova2000AAS..143....1G_CDS}. The motivations and criteria for the inclusion of sources in our compilation are fully discussed in Sect.~\ref{sec:historical-dtypes}, along with the justification of the data types collated into our catalogue.
These include stellar parameters, SEDs, equivalent widths (EW) of emission lines and other spectroscopically derived features relevant to the study of YSOs, infrared (IR) classes related to YSOs' evolutionary stage and information on their X-ray emission - the collated data are also summarised in App.~\ref{app:DBsummary}. We further added value to our catalogue by employing the curated SEDs to homogeneously derive IR classes for $\sim92\%$ of the sources in the catalogue (Sect.~\ref{sec:alphaindex}) and by utilising a multifaceted approach to evaluate and multiplicity of sources (Sect.~\ref{sec:multiplicity}). Finally, in Sect.~\ref{sec:discussion} we discuss the incidence of massive stars and contamination by other types of sources in our catalogue.

\section{Data Curation}\label{sec:data}

\begin{figure}   
    \includegraphics[width=0.9\linewidth]{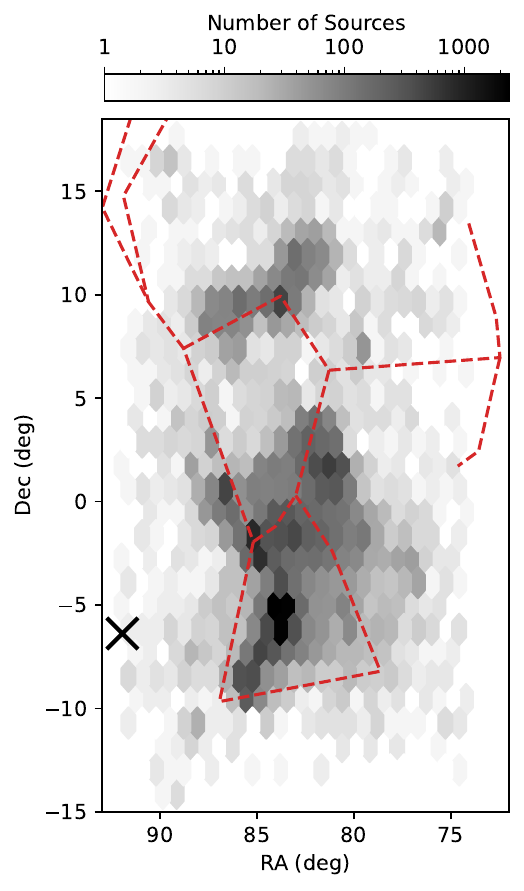}
    \caption{Density distribution of sources included in our data compilation throughout the Orion Star-Formation Complex. The location of Monoceros R2 region - excluded from our compilation - is marked as a black `X'.}
    \label{fig:Orion_FOV}
\end{figure}

\begin{figure*}
    \centering
    \includegraphics[width=0.95\linewidth]{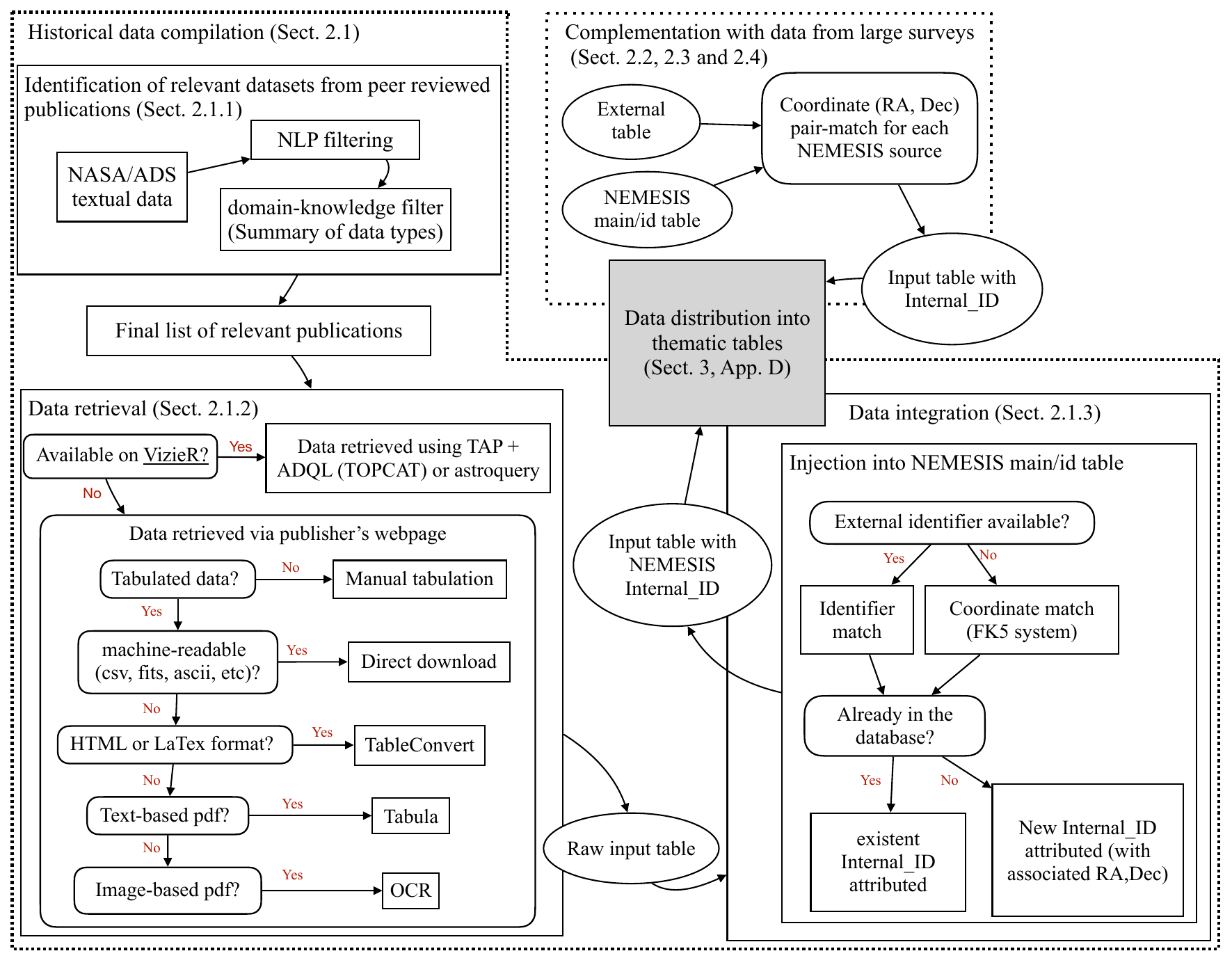}
    \caption{Schematic workflow representation of the data curation process described in Sect.~\ref{sec:data}}
    \label{fig:workflow}
\end{figure*}

We reviewed the published literature targeting OSFC with the goal of building the largest catalogue of candidates and confirmed YSOs in the region. Our definition of the OSFC field encompasses 564 square degrees that cover the box between RA 74.2 and 92 deg and DEC $-14.1$ and $+17.6$ deg. We considered YSOs explicitly listed as members of the Monoceros R2 region (at RA, DEC: 91.94825, -6.37850 deg) to be outside our scope, although some of those were still indirectly included in our compilation. The sky distribution of 27,879 collected YSOs is shown in Fig.~\ref{fig:Orion_FOV}. Fig.~\ref{fig:workflow} presents a schematic representation of the data curation workflow further described in this section.

\subsection{Data mining strategy for historical data}\label{sec:data-historical}

The last comprehensive historical literature review for the OSFC dates back to the Handbook of Star Formation Regions \citep{HandbookStarFormingI_2008hsf1.book.....R,HandbookStarFormingII_2008hsf2.book.....R,Bally2008hsf1.book..459B} - more than 15 years ago. However, that review did not provide a list of known YSOs in the region that could give us a head start.
Hence, instead of starting from previous literature reviews, we focused on the entire body of peer-reviewed publications targeting the region, which have been indexed by the SAO/NASA Astrophysics Data System \citep{Kurtz2000AAS..143...41K}. 
This exercise was carried out with the objective of identifying and data-mining publications that focused on the identification and/or characterisation of YSOs in the OSFC. With our scope on building a census of sources where either a protostar or a pre-main sequence star is already observable, we are, therefore, roughly covering Class 0 to Class III YSOs. Although some cores and Herbig Haro (HH) objects may be collaterally included in our historical data compilation, those sources and others intimately related to the earlier stages of the star formation (like starless cores, jets, globules, etc.) are beyond our scope. 

At the high stellar-mass end, our compilation included both higher-mass YSOs, such as Herbig Ae/Be, as well as O-type stars, which albeit no longer young within their own evolutionary path, are a youth indicator for a coeval population covering the full mass spectrum.
Whenever possible, we have identified and flagged massive stars inferred to be beyond their zero-age main sequence (ZAMS) (see also Sect.~\ref{sec:massivestars}).

\subsubsection{Identification of Relevant Datasets through the NASA/ADS Database} \label{sec:NASA/ADS} 

 The SAO/NASA Astrophysics Data System \citep[NASA/ADS;][]{Kurtz2000AAS..143...41K} is a digital library operated by the Smithsonian Astrophysical Observatory under a NASA grant, which concentrates bibliographic collections for astronomy and physics. We used a Python wrapper, the \citet{ADS_API} \citep{ADS_API_Python}, to query NASA/ADS for bibliographic references that targeted the OSFC. We started with a simple query for \texttt{Orion} to retrieve $\sim 118,000$ bibliographic entries. From there, we restricted our search to the \texttt{Astronomy Database}, excluded \texttt{non-refereed} entries prior to 2020, and refereed entries published before 2018, but never cited. After this initial filtering, we retrieved textual meta-data (titles, abstracts, and keywords) for $\sim 11,000$ bibliographic entries. 

Next, we applied a ``bag-of-words'' method to process these bibliography entries. We started by merging titles, abstracts, and keywords for each bibliographic entry into a single text-based entry. We then employed Natural language processing (NLP) methods to tokenise, lemmatisate, merge multi-word terms into single tokens, remove stop words, and homogenise spelling. The processed text was then scored on the basis of the occurrence of contextual tokens provided in an input vocabulary. This input vocabulary included terms such as \emph{Young Stellar Object}, \emph{Star Formation},  \emph{Young}, \emph{Equivalent Width}, \emph{Emission Line}, \emph{Line Profile}, \emph{Pre Main Sequence}, \emph{Accretion}, \emph{H-alpha}, \emph{T Tauri}, \emph{Herbig}, \emph{disc}, \emph{protoplanetary}, \emph{IR-excess} and \emph{protostar} along with alternative versions of these terms, including their abbreviations, synonyms, and derived words. 

A subset of $\sim 4\,500$ bibliographic entries was highly scored on the basis of this vocabulary. This subset was examined by a human who inspected abstracts and manuscripts, labelled entries by thematics, and ruled out bibliographic entries that were not relevant to our compilation. For example, we excluded purely theoretical studies, investigations of YSOs in other star-forming regions that employed previously published data in Orion as a comparison to their results, studies reusing previously published observational data without relevant added value quantities, publications based on purely qualitative data, studies focused on the interstellar medium or the Orion nebula itself, and studies focused purely on massive stars.

After this initial human-with-domain-knowledge filtering, we reduced our list to 1201 bibliographic entries with observational data deemed relevant to our data curation. Next, this bibliography was carefully read and the data and acquisition methods were summarised as possible. This process allowed us to further discard publications where: (i) the data \emph{actually provided} were not relevant to the NEMESIS project; (ii) the tabulated data were not provided by authors; (iii) the data provided was redundant and introduced in another related publication, (iv) data products were limited to photometric catalogues for the OSFC field without explicit YSO identification, such as the data could be fully retrieved via cone-search using the VizieR-SED tool (see Section \ref{sec:VizierSED}); or, finally, (v) the data were obtained several decades ago and where already included in previous data compilations in a suitable machine-readable format, hence also redundant. This process also included the identification of relevant meta-data within the text of the papers that were worth tabulating.

This phase of our project was finalised in January 2023, although some more recent publications were later identified and manually included in the compilation. The processing of the relevant bibliography also allowed us to identify a dozen other scientific papers that included relevant data for young stars in Orion but did not mention the higher-level keyword ``Orion'' in their title, abstract or keyword list, hence not being identified by our first query with NASA/ADS. 

\subsubsection{Data retrieval}\label{sec:histDataRetrival}

From the 2000s onwards, a growing number of authors made data associated with their peer-reviewed publications at least partially available at CDS via the VizieR catalogue service \citep[hereafter \citeauthor{VizieR};][]{Ochsenbein2000AAS..143...23O}, making data collation very straightforward once the relevant references and tables were identified. However, the data relevant to this compilation were often absent or only partially uploaded to CDS. We therefore further employed one or more of the six strategies detailed below to extract data from scientific publications. These are briefly described in order of efficiency (and priority). 

\begin{enumerate}
    \item Data were available via VizieR, or could be retrieved with widely used tools under table access protocol \citep[TAP][]{Dowler2019ivoa.spec.0927D}, for example using Astronomical Data Query Language \citep[ADQL][]{Osuna2008ivoa.spec.1030O}, TOPCAT \citep[Tool for OPerations on Catalogues And Tables;][]{Taylor2005ASPC..347...29T} or \texttt{astroquery} \citep{Astroquery2019AJ....157...98G}. 
    \item The table was available in electronic format at the publisher's website in some standard data format (\emph{csv}, \emph{ascii}, \emph{tbl}, etc.).
    \item Tables were only available as displayed on the publisher's website in HTML or \LaTeX\ format. \citet{TableConverter} tools to convert them to a suitable machine-readable format.
    \item Tables were only available as published, as part of a text-based \texttt{.pdf} file. The \citeauthor{Tabulalink} \citep{tabula} to convert it into a suitable format. 
    \item Tables were only available as part of an image-based \texttt{.pdf} (common for older publications) or as an image format within the publication. In these cases, a \texttt{.png} or \texttt{.jpeg} image of the table was inputted to a free OCR tool (typically \citeauthor{OCR}). However, we noticed that OCR tools available to us at the time were prone to imprecision affecting decimal numbers and symbols such as `+' and `-'. Hence, this method required a careful visual comparison between the input image table and the output ASCII table to correct for imprecisions. 
    \item Relevant data were not available in tabular format and had to be interpreted and retrieved from the publication's text.  
\end{enumerate}

Tab.~\ref{tab:papers_with_data} summarises the methods for data retrieval employed in each of the 217 scientific publications with data collated into the NEMESIS YSO catalogue for the OSFC.

\subsubsection{Data Integration}\label{sec:data-integration}

We processed the data included in the historical compilation to build a master catalogue in which each YSO-candidate previously identified in the literature is attributed a unique identifier. Sect.~\ref{sec:historical-dtypes} provides more details of this processing. For each new literature reference integrated into our compilation, the new data set was matched to the master catalogue preferentially based on the identifier codes defined in previous studies. When unique identifiers were unavailable, we employed TOPCAT's Pair-Match to join tables based on coordinate matching in the FK5 system. We used a 2" matching radius as the default. After testing different radius to match publications sharing common identifier codes, this was most often the best one. However, in cases where the source publication explicitly described larger PSF or astrometric precision, a larger radius of up to 10" was required.
At each step, unmatched sources were added as new sources and received appropriate indexing. Our unique identifier, namely \verb|NEMESIS_ID|, follows integer numbers from 1 to 27915. 

\subsection{Data from large photometric surveys}\label{sec:largePhotSurveys}

In addition to the compilation of YSO candidates, we have also built a reference photometric database for the OSFC field based on a list of large-scale photometric surveys, including observations at 34 photometric bands and covering the range wavelength range 0.15--160 $\mu m$. 
This allowed us to build a regular baseline for our panchromatic compilation, helping us to understand our historical compilation's detection limits and completeness. Furthermore, by retrieving data directly from these surveys' archives and following their user documentation, we could guarantee that the catalogues were properly cleaned before they were integrated into our database. It also helped verify the reliability of the data retrieved using the VizieR-SED tool (Sect.~\ref{sec:VizierSED}). Unless stated otherwise, we used a 2" matching radius to identify counterparts in our catalogue.

\begin{description}
    \item[GALEX DR6+7:] The Galaxy Evolution Explorer (GALEX) provides UV data in two bands at \emph{FUV} (0.15 $\mu m$) and \emph{NUV} (0.2 $\mu m$). We retrieved data for \emph{DR6+7} \citep{Bianchi2017ApJS..230...24BGALEX} through \emph{VizieR}, cleaned them to remove sources contaminated with artefacts.  

\item[PanSTARRS1 DR2:] Data from the Panoramic Survey Telescope and Rapid Response System DR2 \citep[Pan-STARRS;][]{PS12016arXiv161205560C} was retrieved through the \citeauthor{PS1CASJOBS} 
SQL Server. \emph{PanSTARRS1} covers most of the sky above Declination $-30^\mathrm{o}$ in the bands \emph{g$_\mathrm{PS1}$} (0.48 $\mu m$), \emph{r$_\mathrm{PS1}$} (0.62 $\mu m$), \emph{i$_\mathrm{PS1}$} (0.75 $\mu m$), \emph{z$_\mathrm{PS1}$}(0.87 $\mu m$), and \emph{y$_\mathrm{PS1}$} (0.96 $\mu m$). We retrieved aperture photometric data from the stacked version of the survey and cleaned the dataset as recommended in \citet{PS1_Flewelling2020ApJS..251....7F} to remove sources of poor quality, suspected duplicates, and keep only the best quality detections. 

\item[Gaia DR3:] Gaia provides all-sky photometry in the bands \emph{G$_\mathrm{G}$} (0.50 $\mu m$), \emph{G$_\mathrm{BP}$} (0.59 $\mu m$), \emph{G$_\mathrm{RP}$} (0.77 $\mu m$). We retrieved both photometric and astrometric data for the OSFC field through the \citeauthor{GaiaArchive}. We followed the recommendations in the \emph{Gaia DR3} release papers and 
employed the $C*$ metrics defined by \citet{Riello2021_GDR3} to correct for inconsistency between different passbands. Then, as recommended by \citet{Riello2021_GDR3} and \citet{Fabricius2021}, we limit the effects of brightness excess towards the fainter end of the $G_\mathrm{BP}$ passband by limiting our dataset to stars brighter than $G_\mathrm{BP}=20.9$ mag. Finally, we applied the saturation corrections proposed in appendix C.1 of \citet{Riello2021_GDR3} for the brightest stars.

\item[2MASS PSC:]
Data from the Two Micron All Sky Survey \citep[2MASS][]{Skrutskie2006_2MASS} Point Source Catalogue (PSC) is available all-sky and were retrieved via the NASA/IPAC Infrared Science Archive (\citeauthor{IRSA}). \emph{2MASS/PSC} provides photometry for the near-infrared bands \emph{J$_{2M}$} (1.24 $\mu m$), \emph{H$_{2M}$} (1.66 $\mu m$), and \emph{K$_{s,2M}$} (2.16 $\mu m$). We used the \emph{2MASS} photometric quality flag (\texttt{|ph\_qual}) to clean up the catalogue. Only sources with quality \texttt{A}, \texttt{B} or \texttt{C} were kept in the main catalogue. Low SNR sources (\texttt{ph\_qual}=\texttt{D}) and and upper limit detections (\texttt{ph\_qual}=\texttt{U}) are kept, but as limit values. 

\item[UKIDSS DR9:] Data from the United Kingdom Infrared Telescope (UKIRT) Infrared Deep Sky Survey \citep[UKIDSS;][]{Lawrence2007MNRAS.379.1599L} DR9 were available for part of the OSFC as part of the Galactic Clusters Survey (GCS) and were retrieved through VizieR. The survey includes the infrared bands: \emph{Z$_{U}$} (0.88 $\mu m$), \emph{Y$_{U}$} (1.03 $\mu m$), \emph{J$_{U}$} (1.25 $\mu m$), \emph{H$_{U}$} (1.63 $\mu m$) and \emph{K$_{U}$} (2.20 $\mu m$). We have cleaned the catalogue for duplicates, sources flagged as probable noise. 

\item[UHS DR2:] The UKIRT Hemisphere Survey \citep[UHS;][]{Dye2018MNRAS.473.5113D} covered most of the northern part of the OSFC in the \emph{J$_{U}$} and \emph{K$_{U}$} bands. We have retrieved the data available from DR2 \citep{Bruursema2023AAS...24211808B} through the \citeauthor{WSA} ADQL service. We have cleaned the catalogue for duplicates, sources flagged as saturated or probable noise. 

\item[VHS DR5:] The Vista Hemisphere Survey \citep[][]{McMahon2013Msngr.154...35M} covered the southern part of the OSFC in the \emph{Y$_{V}$} (1.02 $\mu$m), \emph{J$_{V}$} (1.25 $\mu$m), 
and \emph{K$_{s,V}$} (2.14 $\mu$m) bands. We retrieved data from DR5 through VizieR and cleaned for duplicates, sources flagged as probable noise. 

\item[WISE:] The Wide-field Infrared Survey Explorer \citep[WISE][]{Wright2010_WISE} observed the whole sky in four mid-infrared bands: \emph{W1} (3.4 $\mu$m), \emph{W2} (4.6 $\mu$m), \emph{W3} (12$\mu$m), and \emph{W4} (22$\mu$m). For the \emph{W1} and \emph{W2} bands, we collected data from the \emph{CatWISE2020} Catalogue \citep{CatWISEMarocco2021ApJS..253....8M}, which provides photometry extracted from co-added images produced as part of the \emph{unWISE} extension \citep{UnWISE2019ApJS..240...30S} based on all available observations from \emph{WISE} and \emph{NEOWISE} \citep[post-cryogenic reactivation WISE mission][]{NEOWISEMainzer2014ApJ...792...30M}. For the \emph{W3} and \emph{W4} bands, no new data were acquired after the cryogenic mission, and the best data release to date is still the All-Sky Release Source Catalog \citep[AllWISE][]{ALLWISE_catalogue}. We retrieved \emph{AllWISE} data directly from the survey website\footnote{\url{https://wise2.ipac.caltech.edu/docs/release/allwise/}}. As the WISE survey was not originally designed for the specific study of YSOs, the survey's source extraction process was not optimised for the usual sky background in star-forming regions - an issue previously acknowledged and discussed by a number of authors \citep[e.g.][]{Koenig&Leisawitz2014,Marton2019}. The direct use of the \emph{AllWISE} catalogues for studying YSOs is thus prone to include numerous sources with spurious photometric measurements, especially in the two longer wavelengths. In App. \ref{app:allwise}, we discuss how we built a Random Forest classifier trained on labelled image stamps of \emph{AllWISE} \emph{W3} and \emph{W4} bands to identify and remove probable spurious detections from our catalogue. 

\item[SEIP:] Data from the Spitzer Enhanced Imaging Products \citep[SEIP][]{SEIP} were retrieved via IRSA. The SEIP Source List catalogue was downloaded and processed according to the \citet{SEIP_suplement}. This data release included photometry for the four \emph{IRAC} \citep{Fazio2004_IRAC} bands \emph{I1} (3.6 $\mu m$), \emph{I2} (4.5 $\mu m$), \emph{I3} (5.8 $\mu m$), and \emph{I4} (8.0 $\mu m$), and for the \emph{MIPS} \citep{Rieke2004_MIPS} band at \emph{M1} (24$\mu m$). We adopted the photometry for a 3.8" diameter aperture. 
 We noticed that the complete cleaning recommendations in the Explanatory Supplement were too restrictive, especially for regions of the OSFC with an augmented sky background typical of young star-forming regions. Therefore, we limited our cleaning steps to removing extended and saturated sources. 
 
\item[Herschel-PACS PSC 2.0:] We have also collected data from the Herschel-PACS Point Source Catalogue 2.0 \citep{Marton2024AA...688A.203M}, which recently employed a hybrid strategy that combined classical source detection and machine learning techniques to provide an enhanced version of the Herschel/PACS Point Source Catalogue with much higher completeness levels than previous versions. This catalogue provides data in the three PACS bands \emph{P1} (70 $\mu m$), \emph{P2} (100$\mu m$), and \emph{P3} (160$\mu m$). For this specific survey, we employed a 3" matching radius.
\end{description}

\subsection{Data from large spectroscopic surveys}\label{sec:spec-surveys}

We have further complemented our catalogue with stellar parameters derived as part of a series of large-scale spectroscopic surveys briefly summarised here.

\paragraph{Gaia DR3:} Gaia DR3 includes spectroscopic data products based on low-resolution spectra from the Blue and Red Photometers \citep[BP/RP spectra $R\sim30-100$;][]{BPRPspectra2023AA...674A..27A,BPRPspectra2023AA...674A...2D} and medium resolution spectra from the Radial Velocity Spectrometer \citep[RVS; $R\sim 11 500$][]{RVS2023AA...674A..29R}. Along with a series of stellar parameters, the former includes the $H\alpha$ emission line (Sect.~\ref{sec:acc}) with pseudo-equivalent widths estimated as part of the Extended Stellar Parametrizer for Emission-Line Stars (ESP-ELS) \citep{Apsis2023AA...674A..26C}, and the latter includes the Ca II infrared triplet \citep{Lanzafame2023AA...674A..30L}.

\begin{description}

\item[RAVE DR6:]
The Radial Velocity Experiment \citep[RAVE][]{RAVEDR6_Steinmetz2020AJ....160...82S} observed medium-resolution spectra (R $\sim$ 7500) covering the Ca-triplet region. In its Data Release 6 it provides stellar parameters ($\log{g}$, $T_\mathrm{eff}$, metallicity and radial velocity) for 21 sources in our catalogue. 

\item[LAMOST DR10:] The General Survey of the Large Sky Area Multi-Object Spectroscopic Telescope \citep[LAMOST;][]{Zhao2012RAA....12..723Z} has been spectroscopically surveying half of the sky with low- (LRS: $R\sim 1,800$) and medium- (MRS: $R\sim 7,500$) resolution spectra. In its current public release, DR10, LAMOST data products relevant to our compilation in 4 of its released catalogues (\verb|LRS_stellar|, \verb|LRS_astellar|, \verb|LRS_mstellar|, and \verb|MRS_stellar|). We have downloaded the \emph{v2.0} of these DR10 catalogues directly from \citet{LAMOST}. This included LAMOST-derived parameters for 5,180 sources in our catalogue. 

\item[GALAH DR4:] The recent Galactic Archaeology with HERMES Survey Data Release 4
\citep[GALAH DR4][]{Buder2024arXiv240919858B} included parameters ($[Fe/H]$, radial velocity, $\log{g}$, $T_\mathrm{eff}$, $v\sin{i}$, EWs for $H\beta$, $H\alpha$, and Li) from high-resolution spectroscopy ($R\sim28,500$) for 2058 sources in our catalogue. 

\item[Gaia-ESO Survey DR5:] The Gaia-ESO Survey \citep{Gilmore2022AA...666A.120G, Randich2022AA...666A.121R} DR5 5.1 Catalogue \citep{Hourihane2023AA...676A.129H} provides stellar parameters and a series of emission lines derived from spectra observed with ESO's UVES and GIRAFFE instruments and includes data for 741 sources in our compilation. 

\item[APOGEE DR17:] Although results from a dedicated sub-survey of APOGEE in Orion are included in our historical compilation \citep{2018ApJS..236...27C}, we have further complemented our catalogue with stellar parameters derived with their ASPCAP pipeline in the context of APOGEE DR17 \citep{APOGEEdr172022ApJS..259...35A}, including stellar parameters for 9,429 sources in our compilation. 

\item[ASCC-2.5 V3:] The All-sky Compiled Catalogue of 2.5 million stars, 3rd version \citep{2001KFNT...17..409K} which provides a large compilation of spectral types for bright stars and includes spectral types for 1,069 sources in our catalogue. 
\end{description}

\subsection{VizieR Photometry Viewer}\label{sec:VizierSED}

We have further complemented our database with photometric data collected using the \citeauthor[hereafter VizieR-SED][]{VizierSED}.
Powered by CDS, VizieR-SED 
is a tool intended for the visualisation of photometry around a given sky position. Hence, for an input coordinate or source name, it returns photometric observations ``extracted around a sky position from photometry-enabled catalogues in VizieR'' \citep[][]{VizierSED}, where this photometry data can be exported in typical flux density units. The interpretation of the tool's output as an SED is cautioned as CDS provides ``no guarantee that all photometry points correspond to the target, especially for extended sources'' \citep*{VizierSED}\footnote{As part of our experience with VizieR-SED we second this recommendation and strongly advise against the indiscriminate use of SEDs retrieved with the tool for scientific purposes that require precise SEDs. One of the issues we identified can be traced back to the zero points adopted for magnitude-flux conversions in the tools' back-end. The zero point values used can be recovered from the VizieR meta data tables \texttt{METAfltr} and \texttt{METAphot} at VizieR's \texttt{ReferenceDirectory}. For example, 2MASS zero points adopted by CDS can be traced back to  \citet{2010PASP..122.1437P}, and are offset in relation to official zero point values recommended by the survey's documentation \citep{2003AJ....126.1090C}, yielding a $\sim 3\%$ offset in the flux densities converted by VizieR.}. Nevertheless, VizieR-SED provides a powerful tool for data mining the CDS database, and we have employed it to further search for relevant archival photometry for our sources. For that, we have followed their API's recommendation to develop a Python script to query VizieR around the coordinates of each of our YSO candidates. We initially used a search radius of 5". This wider search radius was chosen purposely, as it is prone to including contaminants from neighbouring sources, and we have taken advantage of this to flag sources likely susceptible to such contamination (see Sect. \ref{sec:bin:bigdata}). 

For the specific purpose of complementing the SEDs of sources in our catalogue (Sect.~\ref{sec:SED}), we have reduced this radius to 2" to minimise contamination by neighbouring sources. We visually inspected a large number of SEDs produced with VizieR-SED data and compared them with SEDs built using photometry from our historical compilation (Sect.~\ref{sec:data-historical}) and large photometric surveys (Sect.~\ref{sec:largePhotSurveys}), which had their magnitude-flux conversions carefully carried out by ourselves using meta-data available in their original publication. This procedure allowed us to identify and exclude from the SEDs data from surveys showing large systematics compared to the larger body of data available. We have post-processed VizieR-SED outputs to remove data coming from all CDS tables already included in the compilation (\emph{i.e.,} from Sect.~ \ref{sec:data-historical} and \ref{sec:largePhotSurveys}). 

We also noticed large spreads in flux density in the SEDs generated by VizieR-SED, especially in the optical range 0.4-0.8 $\mu$m. By further inspecting these data original tables, we verified that although a fraction of the spread remained unexplained, most of these could be traced back to either multi epoch surveys  (YSOs are typically photometric variables as discussed in Sect.~\ref{sec:var}) or to surveys targeting extended sources and publishing photometry extracted for different aperture sizes (e.g. SDSS). Finally, we opted to avoid this wavelength range and kept only data in the ranges 0.1-0.4 $\mu$m and 0.8-1000$\mu$m, which seemed less affected by these issues.

We have also identified a significant number of duplicated data, introduced by scientific publications re-reporting photometric data from large surveys. For example, although we have removed all 2MASS official data release tables - as this data has already been included in our compilation in Sect.~\ref{sec:largePhotSurveys} - many sources had VizieR SEDs with dozens of 2MASS JHKs data points, most of which having the same flux values but originating from different CDS tables. Together with multi-epoch data, this type of duplicate is problematic for quantities derived from SED-fits, as in the added-value quantities discussed in Sect.~\ref{sec:alphaindex}, as it biases the fit results towards the filters with a larger number of photometric points. We have, thus, further processed the SEDs from VizieR-SED to aggregate photometric measurements reported under the same filter name by averaging their fluxes weighted by their uncertainties and propagating these uncertainties accordingly. 

\section{Criteria for inclusion in this compilation and description of historical data-types}\label{sec:historical-dtypes}

Throughout our historical data curation (Sect.~\ref{sec:data-historical}), we focused on sources in the OSFC field previously reported as young in the literature. We have adopted a positive evidence rule for inclusion in the catalogue, where a given source had to be reported as a YSO candidate by any but at least one of the types of study discussed in this section. Our approach thus prioritises completeness over purity. In this section, we summarise the main YSO identification approaches employed by the scientific publications included in our historical compilation. We note that a revision of the different criteria adopted in the literature was beyond our scope. However, we were still interested in collating data that could support the ranking sources as \emph{bona fide} YSOs. Hence, in this section, we also discuss a series of data types collected for this purpose. The various data types included in the catalogue are also summarised in Apps.~\ref{app:historical_bib} and \ref{app:DBsummary}.

\subsection{Colour-magnitude and Hertzsprung-Russell diagrams}\label{sec:hrdiagram}

Newly formed stars are still contracting gravitationally and acquiring part of their final mass from their surroundings. Because they are still larger and colder than their Main Sequence (MS) counterparts, YSOs are located above ZAMS when placed in colour-magnitude (CMD) or Hertzsprung-Russell diagrams (HRD). Our compilation includes all sources reported in the literature as candidate YSOs from CMD selections. \citet{2004ApJ...610.1064B} estimate that optical-near-IR (OIR) CMD YSO selections based solely on photometry suffer at least $\sim 25\%$ contamination by field stars. This contamination level can be greatly minimised when spectroscopic constraints for stellar parameters are available, enabling reliable transposition from observed CMDs into HRDs. Although we included all CMD-selected YSO candidates in this compilation, we also collated products from spectroscopic surveys that can help address the purity of our YSO candidate list. These data products are summarised in Tab.~\ref{tab:hr_diagram} and discussed in the following sections.

\subsubsection{Spectral Types}\label{sec:SpT}

Our historical compilation yielded spectral types for 7,485 sources. With the complementation by large surveys in Sect.~\ref{sec:spec-surveys}, a total of 11,497 sources have spectral types. We tabulated these spectral types in the MK system, with luminosity class, or peculiar spectral characteristics, provided as an extra flag when available. When multiple previous studies provided independent spectral type derivations, a list of values is reported. 

\subsubsection{Effective Temperatures}\label{sec:Teff}

Our compilation included 73,808 measurements of effective temperature ($T_\mathrm{eff}$) for 17,589 sources collated from 42 publications. These measurements have been flagged according to the method used for derivation, where: \texttt{SpT-PHOT} indicates $T_\mathrm{eff}$ derivations based on standard tables of combined spectral-type and photometric data \citep[e.g.,][]{PecautMamajek2013}; \texttt{SED} indicates derivations based on SED-fitting to models \citep[e.g.][]{2008AandA...492..277B}; \texttt{SPEC} indicates derivations based on fitting of observed spectra to spectra libraries \citep[][]{2018AJ....156...84K}; \texttt{PHOT} indicates derivations based on fit to photometric data using Markov Chain Monte Carlo or Bayesian methods \citep[e.g.][]{2012ApJ...748...14D}. We did not include $T_\mathrm{eff}$ estimations based solely on CMD placement or colour/magnitude-$T_\mathrm{eff}$ conversions. 

\subsubsection{Bolometric values}\label{sec:Tbol}

CMD YSO selection is only possible for less embedded sources that have a significant portion of their radiation already visible. For more embedded sources, a more useful parameter space is the Bolometric luminosity-temperature (BLT) diagram \citep[e.g.,][]{Chen1995}. Hence, we also collected 700 derivations of bolometric temperatures ($T_\mathrm{bol}$) for 381 sources derived based on SED-fitting methods in 3 previous studies. 

\subsubsection{Attributes not included}

The derivation of luminosities requires knowledge of the distance to the sources and the amount of interstellar extinction in its line of sight. Luminosities estimated prior to Gaia often carry large uncertainties as a result of the uncertainties behind distance estimations. Moreover, extinction estimations are often degenerate with the $T_\mathrm{eff}$ derivations. Thus, we opt to leave these two quantities out of our data collection because they are often biased by the type of data and methods available at the specific time of publication. 

\subsection{Spectral Energy distributions}\label{sec:SED}

\begin{figure}[htb]
    \centering
    \includegraphics[width=0.9\linewidth]{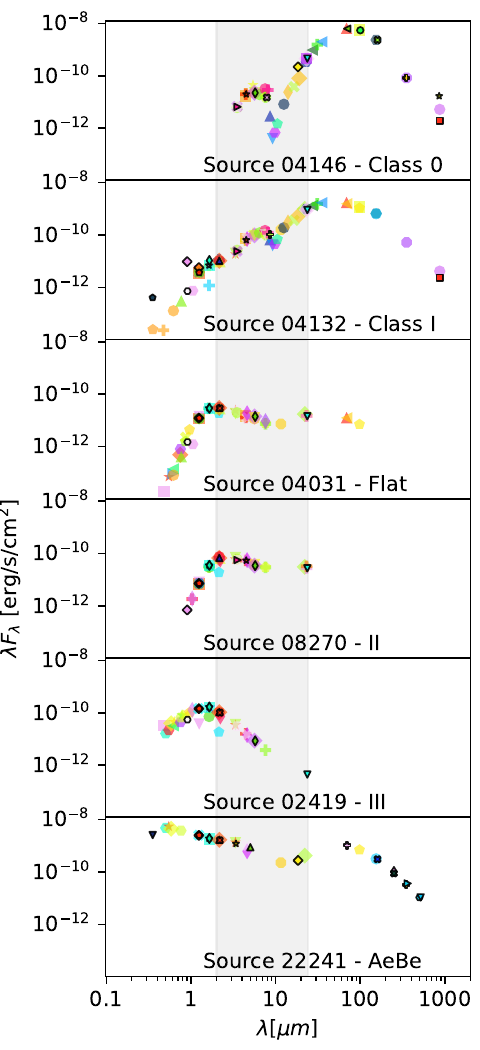}
    \caption{Examples of SEDs from our database for different types of YSOs \. From top to bottom: a Class 0, a Class I, a Flat-Spectrum, a Class II, a Class III, and a Herbig AeBe star. See YSO classes discussion in Sect.~\ref{sec:ir_emission}. Photometric data collated by us is plotted in coloured symbols, while data points retrieved by processing data from the VizieR-SED tool are plotted contoured in black. The grey area shows the wavelength range used for deriving $\alpha_{IR}$ indices in Sect.~\ref{sec:alphaindex}.}
    \label{fig:SED-prototypes}
\end{figure}

As part of our historical compilation, we have collated photometric data covering the 0.1-1000 $\mu$m range and processed it into SEDs. When these data were provided as magnitudes, their photometric systems were identified from their publications, and magnitudes were converted into flux densities. We have preferentially adopted filter profile information from the Spanish VO Filter Profile Service \citep[SVO-FPS;][]{SVO2012ivoa.rept.1015R,SVO2020sea..confE.182R}. When photometric system information could not be recovered from the publication or the telescope/instrument webpages or official publications, we adopted generic values in the Johnson-Cousin systems. A significant number of photometric tables out there include data re-reported from previous studies. We attempted to reduce duplicates by comparing magnitudes in the same filter collected for each source down to a precision of 0.01 and merging duplicates. In the present version of our database, we have not included limit values. For data related to the large photometric surveys described in Sect.~\ref{sec:largePhotSurveys} we have discarded data collected in the historical compilation and instead introduced our own processing of the photometry of these surveys. This greatly complemented the SEDs and guaranteed that $\sim 94\%$ of sources in our catalogue had at least 5 data points in their SEDs with an average of 22 datapoints per SED. Examples of the SEDs collected are shown in Fig.~\ref{fig:SED-prototypes}. 

\begin{figure}
    \centering
    \includegraphics[width=0.9\linewidth]{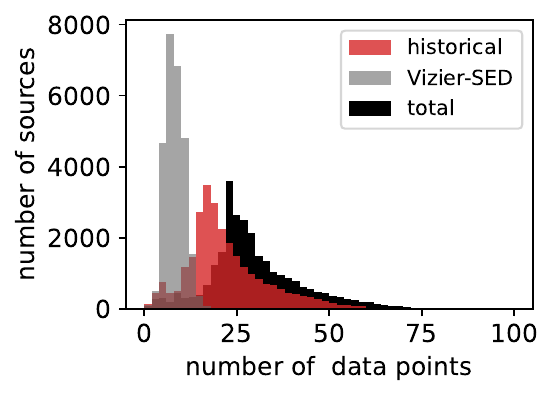}
    \includegraphics[width=0.9\linewidth]{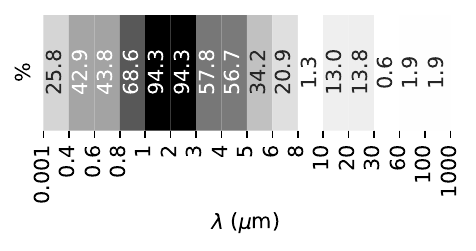}    
    \caption{\emph{Top:} Number of photometric data points in the SEDs collated into the NEMESIS YSO catalogue for the OSFC. \emph{Bottom:} Percentage of SEDs containing data points at a given wavelength range.}
    \label{fig:SED-number}
\end{figure}

Next, we employed data obtained with the VizieR-SED tool to further complement the SEDs. As described in Sect.~\ref{sec:VizierSED}, these data have been post-processed to obtain averaged values. Data from this complement is also shown in Fig.~\ref{fig:SED-prototypes}, and this complementation helped increase the average number of data points per SED to 29, with 96\% of SEDs having at least 5 data points. The typical number of points per SED and their incidence per wavelength range are illustrated in Fig.~\ref{fig:SED-number}. A general description of the SED database is provided in Tab.~\ref{tab:SED}. Additionally, flags with the counts of SED data-points per wavelength range are also provided and are aimed at facilitating the usage of the catalogue.

\subsection{Infrared emission by circumstellar material}\label{sec:ir_emission}

At the earliest phases of young stellar evolution, the dusty envelope surrounding YSOs obscures their radiation in the optical wavelengths, which is reemitted at longer infrared and radio wavelengths. Various YSO identification and characterisation methods stem from studying this reemitted radiation. Following the seminal work of \citet{Lada:87a}, one of such methods is based on the derivation of the IR spectral index, $\alpha_{\mathrm{IR}}$, which gauges the amount of circumstellar material as the shape of the SEDs of YSOs in the IR ($\lambda\geq 2~\mu$m):

\begin{equation}\label{eq:alpha}
  \alpha_{\mathrm{IR}}=\frac{d\log{(\lambda F_\lambda)}}{d\log{\lambda}}.
\end{equation}

\noindent\citet{Lada:87a} initially assigned YSOs into three classes associated with the evolution of the forming stars from core collapse to a pre-main sequence (PMS) phase: Class I sources were protostars still embedded in their natal material, Class II were disc-bearing PMS stars (often associated with accreting T Tauri stars), and Class III sources were PMS stars where the dusty circumstellar disc responsible for the IR-excess of YSOs has mostly dissipated. Later, Class 0 sources were included to account for the youngest deeply embedded YSOs \citep{Andre:93a}, and the flat-spectrum class described sources with characteristics between Classes I and II \citep{Greene1994}. Examples of SEDs of those standard IR classes are shown in Fig.~\ref{fig:SED-prototypes}. With the advent of large-scale IR facilities (\emph{e.g.,} \emph{Spitzer}: \citet{Werner2004ApJS..154....1W_spitzer}, \emph{WISE} \citep{Wright2010_WISE} and \emph{Herschel} \citep{Pilbratt2010AA...518L...1P_Herschel}), the increase and diversification of the available IR data have helped uncover physical aspects of the evolution of the YSO that could no longer be fully explained within \citeauthor{Lada:87a}'s standard classification scheme, leading to the proposal of alternative classification schemes that include classes such as Transition discs \citep[e.g.][]{2007ApJ...671.1784H,Luhman2008,2009AandA...504..461F,McClure2010,Dunham2014prpl.conf..195D,2016ApJS..226....8K,2013ApJS..207....5F,Espaillat2014,2018ApJ...863...13G}. 

Rather than looking at SEDs, a range of IR classification schemes combines selection cuts within sets of CMDs and colour-colour diagrams into decision tree models to isolate the loci occupied by YSOs in these diagrams \citep{Gutermuth2009,2012AJ....144...31K, 2012AJ....144..192M,2012AJ....144...31K,2007ApJ...671.1784H,2009ApJ...707..705H,2010ApJ...722.1226H,2014ApJ...794...36H,2015AJ....150..100K,2018ApJS..236...27C}. These approaches cover the initial identification phase for a substantial portion of the YSO candidates featured in our data set. Although these techniques include cuts to exclude sources with colours and magnitudes consistent with active galactic nuclei (AGN), Star-Forming Galaxies, sources from the asymptotic giant branch (AGB), polycyclic aromatic hydrocarbon emission and knots of shock-heated emission, contamination by these sources is expected and further addressed in App.~\ref{app:galaxy}. 

All sources reported by studies applying infrared classification schemes to study YSOs in the OSFC field were included in our historical compilation. More details on the data types compiled are given in the next section and in Tab.~\ref{tab:app:disc}. Although we have not compiled infrared excess measurements, we included a flag pointing to bibliographic references that include these attributes in the database.

\subsubsection{Standard IR classification}\label{sec:alphaindex}

\begin{figure}
    \centering
    \includegraphics[width=0.9\linewidth]{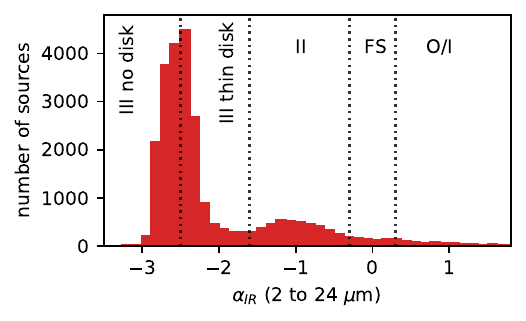}
    \caption{Distribution IR spectral indices, $\alpha_{IR}$, estimated for sources in the NEMESIS Catalogue of YSOs in the OSFC using all photometric data available in the wavelength range 2--24$\mu$m. Dotted lines show the limits between classes (Tab.~\ref{tab:yso_classes}).}
    \label{fig:alpha-index}
\end{figure}

\begin{figure}
    \centering
    \includegraphics[width=1\linewidth]{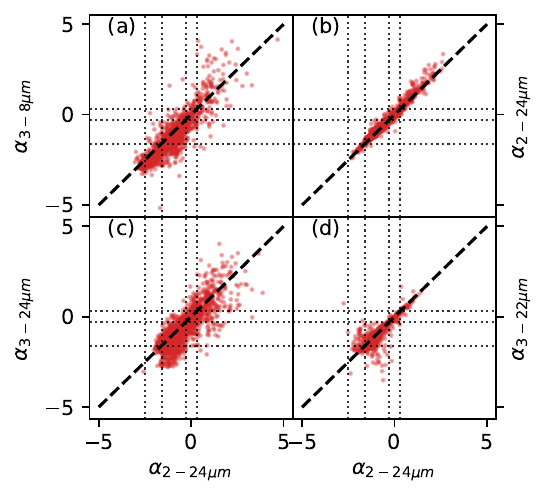}
    \caption{Comparison of $\alpha_{IR}$ indices derived in this study (Fig.~\ref{fig:alpha-index}) with literature values from: (a) \citet{2015AJ....150..100K}, (b) \citet{2019AandA...622A.149G}, (c) \citet{2012AJ....144..192M}, (d) \citet{2017ApJS..229...28G}
   . Dashed lines reflect a 1:1 equivalence. Dotted-black lines show the limit between YSO classes adopted in this study (Tab.~\ref{tab:yso_classes}). \label{fig:alpha_index_comparison}}
\end{figure}

Because a reported YSO class in any given IR classification was a criterion for including sources in this compilation, we have accordingly also collected such classes as part of our historical compilation. Nevertheless, the evergoing evolution of infrared classification schemes makes the collation and homogenisation of infrared classes derived by different authors an unpractical task. Instead, we employed the SED database described in Sect.~\ref{sec:SED} to homogeneously infer IR classes in the standard classification scheme. For that, we used the procedure described in \citep{NEMESIS_David} to fit Eq. (\ref{eq:alpha}) to SEDs and derive $\alpha_{IR}$ for each source using all data available within 2--24$\mu$m. We were able to derive $\alpha_{IR}$ values for 25\,799 sources (Fig.~\ref{fig:alpha-index}), with the requirement that at least two data points at least 2 $\mu$m afar in wavelength existed within the 2--24$\mu$m range. For assigning IR classes, we have followed the IR class definitions as employed by \citet{2019AandA...622A.149G}, which is outlined in Tab.~\ref{tab:yso_classes}.  Beyond this added value, we have compiled 16\,716 IR classes for 5\,379 sources from 30 publications. Fig.~\ref{fig:alpha_index_comparison} shows a comparison between our own $\alpha_{IR}$-index values and the 4 largest samples of $\alpha_{IR}$-index values we found in our historical compilation. 

\begin{table}
  \caption{Standard YSO IR classes definition adopted in this study , based on $\alpha_{IR}$ indices estimated for the 2--24$\mu$m 
 wavelength range.}%
  \label{tab:yso_classes}               
  \centering                            
  \begin{tabular}{l c}                  
    \hline\hline                        
    Class         & \(\alpha_\mathrm{IR}\) range                                       \\    
    \hline                              
    0/I           & \(\hphantom{{-}0.3 \geq\;}\alpha_\mathrm{IR} > \hphantom{{-}}0.3\) \\         %
    flat spectrum & \(\hphantom{{-}}0.3 \geq \alpha_\mathrm{IR} > -0.3\)               \\
    II            & \(-0.3 \geq \alpha_\mathrm{IR} > -1.6\)                            \\
    III thin disc & \(-1.6 \geq \alpha_\mathrm{IR} > -2.5\)                            \\
    III no disc   & \(-2.5 \geq \alpha_\mathrm{IR}\hphantom{\;>{-}2.5}\)               \\
    \hline                              
  \end{tabular}
\end{table}

Fig.~\ref{fig:alpha_index_comparison} illustrates that data availability as a function of wavelength and methodological differences in the derivation of $\alpha_{IR}$ can affect the values derived, where the spreads observed can be traced back to distinct wavelength ranges employed: 3--8$\mu$m from \citet{2015AJ....150..100K} 2--24$\mu$m from \citet{2019AandA...622A.149G}, 3--24$\mu$m from \citet{2012AJ....144..192M}, and 3--22$\mu$m from \citet{2017ApJS..229...28G}, whereas \citet{2019AandA...622A.149G} shows the closest equivalence to our values by covering a range similar to ours but with different specific data available.  We stress that our $\alpha_{IR}$ indices were derived using all data in the wavelength range $2-24\mu$m \emph{as available}. We required a minimum of two data points at least 2 $\mu$m away in wavelength for $\alpha_{IR}$ derivation. As a result of this requirement, $\sim33\%$ reported $\alpha_{IR}$ values were fitted using only data in the range $2-5\mu$m. As further discussed in App.~\ref{app:alpha_ir}, this reduced data availability in the mid/far infrared may result in an excess of sources miss-identified as having a thin disc. To avoid such bias, we recommend using our $\alpha_{IR}$ in light of the provided data availability flags. In addition to the flags for SED sampling as a function of wavelength from Sect.~\ref{sec:SED}, in Tab.~\ref{app:alpha_ir} we also provide flags with the maximum and minimum wavelength available for $\alpha_{IR}$ derivation.

\subsection{Lithium Depletion}\label{sec:lithium}

Lithium is quickly depleted when parts of the stellar interior reach temperatures of $\sim 2.5$ MK \citep{Soderblom2010}. The location within the stellar interior where this takes place depends on the stellar internal structure and energy transport mechanism. In Solar-type stars, this will happen at the base of the convective zone. In the most massive fully convective stars, this takes place in their core,  and in brown dwarfs below $\sim 0.06 M_\odot$, the core will never get hot enough to destroy its Li \citep{Soderblom2010}. The timescale for Li depletion is estimated between 10-50 Myr for M-type stars, 20 and a few Myr for K-type stars, but it can be much longer than YSO evolution timescales for F and G-type stars \citep{Jeffries2014}. Nevertheless, theoretical investigations suggest that episodic accretion in young stars can produce young stellar objects with significantly augmented central temperatures compared to non-accreting counterparts at the same mass and age \citep{BaraffeChabrier2010}. This effect can severely enhance lithium depletion in accreting YSO. Therefore, lithium detection is an excellent spectroscopic proxy for confirming the youth of YSO candidates. All sources towards the field of the OSFC identified as young through their lithium abundance evidence were included in our catalogue. Lithium observables were available as the EWs of the Li I ($\lambda$6708\AA) line or as Lithium Abundances, $A(\mathrm{Li})$. 9,185 measurements were collated from 31 publications for 6,384 sources. More details on the data types collected are provided in Tab.~\ref{tab:Lithium}.

\subsection{Gravity}\label{sec:gravity}

\begin{figure}
    \centering
    \includegraphics[width=0.8\linewidth]{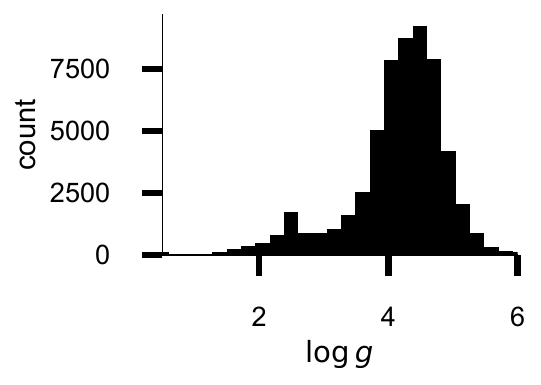}
    \includegraphics[width=0.9\linewidth]{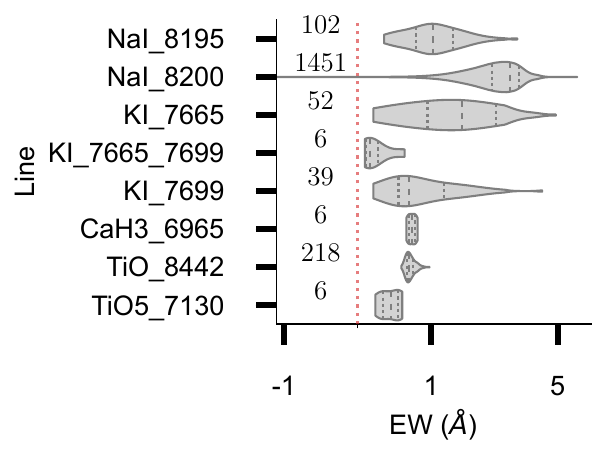}
    \caption{\emph{Top:} Distribution of $\log{g}$ values collated into the NEMESIS Catalogue of YSOs in the OSFC (Sect.~\ref{sec:gravity}). \emph{Bottom:} General distribution of EW values for absorption lines collected as gravity proxy. The dotted red line reflects the continuum level. Violins are shown at a fixed width for improving visualisation. Violin's internal dotted lines reflect distributions' 1st and 3rd quartiles, and dashed lines reflect their median.}
    \label{fig:logg}
\end{figure}

 YSOs are still contracting onto the MS and have weaker surface gravity in comparison to MS stars. For comparison, while YSOs have $\log{g}\sim 3-4.5$, main sequence stars have $\log{g}\sim 3.5-5$, and K\&M-type giants have $\log{g}\sim 1-3$ \citep[with MESA stellar evolution models as reference;][]{Dotter2016ApJS..222....8D}. Gravity observables are thus a valuable asset in distinguishing YSOs against older sources with similar spectral type.
 Measurements of $\log{g}$ are typically obtained by fitting grids of synthetic spectra to observed spectra and are widely available \citep{2016ApJ...818...59D,2018AJ....156...84K,2018ApJ...869...72Y,2020MNRAS.496.4701J,2021MNRAS.506.4232K}. In fact, more than 57k $\log{g}$ measurements were collated from 13 publications, with 12,313 sources having at least one such measurement. We note that Fig.~\ref{fig:logg} points to a degree of contamination by giants in our catalogue. This type of contamination is generally expected in surveys selecting YSOs purely from CMDs or colour-colour diagrams (as in Sect.~\ref{sec:hrdiagram} and Sect.~\ref{sec:ir_emission}). We further evaluate this contamination level in Sect.~\ref{sec:giants} and App.~\ref{app:galaxy}. 

Although widely available, $\log{g}$ derivations require medium to high-resolution spectra. A cheaper alternative is the measurement of equivalent widths of absorption lines known as surface gravity tracers \citep[e.g.,][]{Reid1995AJ....110.1838R,Slesnick2006AJ....131.3016S,Schlieder2012}. For example, the equivalent widths of the Na I doublet lines are expected to be 2 to 3 times stronger in cool M-dwarfs than in PMS stars younger than about 10 Myr \citep{Schiavon1997}. Among many atomic and molecular bands identified as surface gravity tracers, EW observations of the lines Na I, K I, TiO, and CaH 3 have previously been reported for YSOs in the OSFC. We collated 1,880 gravity-related EW measurements from 9 bibliographic references, with measurements available for 1,439 sources. Hence 13,065 sources (47\% of our catalogue) had gravity-related quantities collected. The general distribution of the gravity observables is shown in Fig.~\ref{fig:logg} and data types are summarised in Tab.~\ref{app:tab:gravity}.

\subsection{Spectroscopic signatures of accretion 
and outflows}\label{sec:acc}

Emission lines are a defining characteristic of accreting YSOs \citep{Joy1945,Herbig1962}. These lines are typically formed in outflows or infalling magnetic flows, being intimately related to the accretion process \citep[][]{Hamann1994,Hartigan1995ApJ...452..736H,Hartmann1994,Muzerolle1998ApJ...492..743M_model}. Extensive previous literature, including observational results and predictions from radiative transfer modelling, establishes the relationship between certain emission lines and the phenomenology behind the mass-accretion process in YSOs \citep[e.g.,][]{HamannPersson1992_TTauri,Muzerolle1998AJ....116..455Mobservation,Muzerolle2001obs,Bouvier2007,Lima2010,Kurosawa2011,Natta2014AA...569A...5N,Alcaca2014AA...561A...2A,Alcaca2017AA...600A..20A,Wilson2022}. 

\begin{figure}
    \centering
\includegraphics[width=0.85\linewidth]{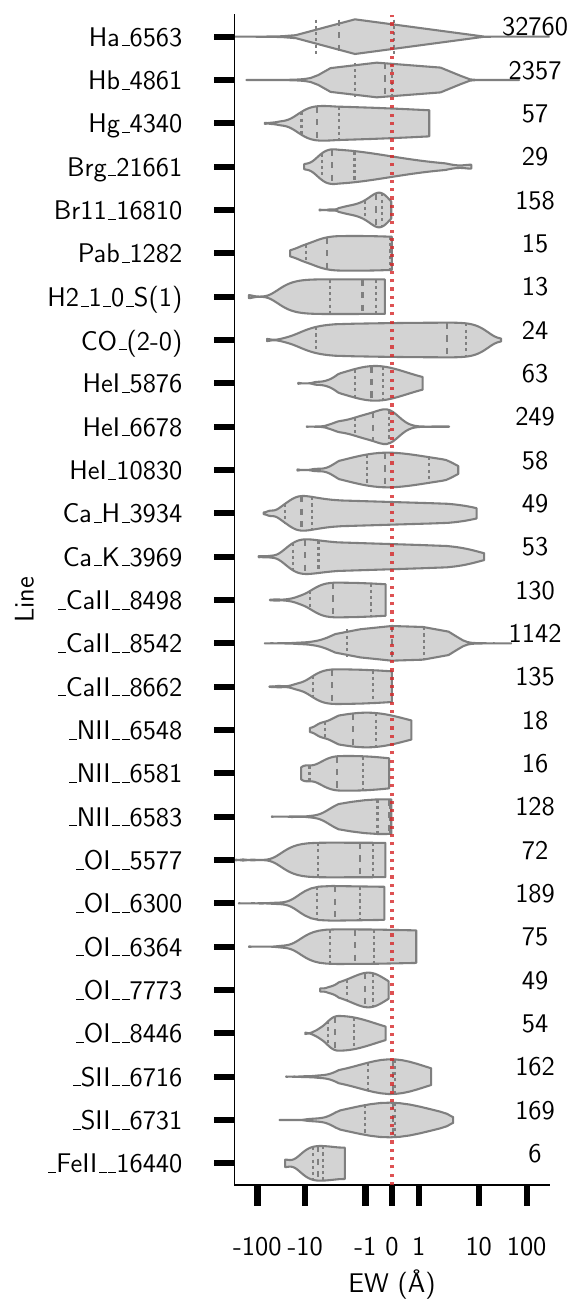}
    \caption{Distribution of equivalent widths for lines related to material inflow, outflow and accretion in YSOs (Sect.~\ref{sec:acc}). The dotted red line reflects the continuum level. Violins are shown at a fixed width for improving visualisation. Violin's internal dotted lines reflect distributions' 1st and 3rd quartile, and dashed lines reflect their median. }
    \label{fig:EW_acc}
\end{figure}

From an observer perspective, in addition to helping identify young stars, measurements of equivalent widths of accretion-related emission lines allow quantifying the flux in excess due to accretion, enabling estimation of mass accretion rates \citep[e.g.][]{2004Natta,2008AandA...488..167S,2009AandA...504..461F,Alcaca2017AA...600A..20A} and help identify \emph{bona fide} YSOs. With a best-effort approach, we collated a large number of measurements of relevant EW of emission lines. Relevant spectral features are briefly discussed in this section if they were available for stars in the OSFC and included in this compilation. Although estimates of mass-accretion rates were also often available, these are heavily dependent on assumptions such as the source radius and distance, and were therefore excluded from this compilation. A summary of EW of emission lines collated can be found in Tab.~\ref{tab:acc} and their distribution is illustrated in Fig.~\ref{fig:EW_acc}.

\subsubsection{Hydrogen Lines}\label{sec:hydrogen_lines}

The strongest emission line observed in the spectra of optically visible YSOs is the H$\alpha$ line \citep[e.g.][]{WhiteBasri2003}. Although broad H$\alpha$ is often associated with the accretion process, narrower H$\alpha$ emission can also be attributed to the chromospheric activity of young stars \citep{Martin1998}. Hence, the association of H$\alpha$ emission with the accretion process requires a threshold for distinction from chromospheric activity. Historically, many authors have applied thresholds with EW(H$\alpha$) at around 10 or 20 \AA$\;$ to identify Classical T Tauri stars (CTTS) and down to 5 \AA$\;$ to identify weak-line T Tauri stars (WTTS). As the diversity of the YSOs studied spectroscopically has grown over time, spectral type-dependent classification criteria were proposed \citep[e.g.][]{Martin1998}. Currently, the most widely used criteria are the ones of \citet{WhiteBasri2003} and \citet{BarradoYNacascuesMartin2003}. However, the increasing availability of large samples of EW(H$\alpha$) seems to indicate that spectral type dependent criteria for discerning CTTS from WTTS fail to identify low-accreting T Tauri stars at the end of their star-disc interaction phase \citep[e.g.][]{2019AJ....157...85B}. This ever-going evolution of classification schemes also means that the labels for actively accreting CTTS and non-accreting WTTS provided in the literature depend on the classification scheme in use at the time of their publication. Therefore, we suppressed these labels from our compilation and focused on measurements of the H$\alpha$ line. Alternatively to EW, some authors focused on the full width at 10\% of the peak of the H$\alpha$ emission profile, $W_{10\%}(H\alpha)$, which is proposed as an accretion diagnosis unbiased by the stellar spectral type \citep{WhiteBasri2003}. Although much less widely available, we also collated these quantities. Measurements for the H$\alpha$ line, whether it is EW(H$\alpha$) or $W_{10\%}(H\alpha)$, are one of the most widely available spectroscopic data products, with 37,449 measurements collated from 49 publications for 21,244 sources - $\sim 76\%$ of the sample had at least one H$\alpha$ measurement.

In the absence of optical spectroscopy around the H$\alpha$ line, other hydrogen lines can be used to constrain accretion. We have collated 2,644 EWs from 10 bibliographic references for 2,501 sources, including observations for other Balmer series lines (H$\beta$, H$\gamma$, and H11), lines from the Paschen series (Pa$\beta$ and Pa$\gamma$) 
and Brackett series (Br$\gamma$, Br11). Finally, the emission of molecular hydrogen, H$_2$, is often interpreted as a signal for strong, bipolar jets or wind and can offer insight into Class 0 outflow activity \citep[e.g.][]{2021ApJ...921..110L}.

\subsubsection{Other lines}\label{sec:otherEW}

Beyond hydrogen lines, various emission lines in the spectra of YSOs can be traced back to inflow and outflow of material inherent to the star-disc interaction and accretion process in these objects. We have collated 2,849 EWs from 16 publications for 1,485 sources in our catalogue which are related to these processes. The general motivations for their inclusion in the compilation are discussed in the remainder of this section.

He I lines are a tracer of outflow phenomena and accretion in YSOs \citep{1995AandAT....8..249K,2008AandA...488..167S}. The He I line profile is sensitive to the kinematics of the stellar wind \citep{2010AJ....140.1214C}. Hence, under certain conditions, He I can be related to chromospheric rather than accretion \citep[e.g.][]{Dupree1992ApJ...387L..85D}. Nevertheless, He I has been suggested as a better tracer for identifying low-accreting stars that would otherwise be deemed non-accreting based on traditional classification schemes based solely on the emission of the H$\alpha$ line \citep{2022AJ....163...74T}. 

Ca II near-IR triplet in emission is also characteristic of classical T Tauri stars \citep{Hamann1994,1998AJ....116.1816H}. In the youngest YSOs, Ca II is likely produced in the protostellar magnetospheric infall of gas \citep{Muzerolle1998ApJ...492..743M_model} and it is considered a common emission feature for Class I YSOs \citep{Azevedo2006,2010AJ....140.1214C}, having previously been used in association with Na I and CO to distinguish them from Class II and III sources \citep{GreeneLada1996}. Although Ca lines in emission are often associated with accretion, these lines are instead observed in absorption for low-gravity sources \citep[e.g.,][]{Reid1995AJ....110.1838R}.

Finally, both forbidden lines, such as $[\mathrm{N} II]$, $[\mathrm{S} II]$, $[\mathrm{Fe} II]$, and $[\mathrm{O} I]$, and some permitted lines (e.g. OI) are also often used as a tracer of outflow material \citep{Hartigan1995ApJ...452..736H,Muzerolle1998AJ....116..455Mobservation} and have been included in the compilation when available.

\subsubsection{Veiling}\label{sec:veiling}

Optical and IR veiling is also a predominant characteristic of actively creating YSOs \citep[][]{Herbig1962,BasriBertout1989ApJ...341..340B}. Veiling is thought to arise from an excess flux originating from high-temperature material in the inner disc regions, the accretion funnel, and the accretion shock regions, which is responsible for diluting the photospheric spectral lines while enhancing the spectral continuum. Veiling at a given wavelength, $r_\lambda$, is typically quantified from a YSO spectrum as the creation between the excess flux to the stellar photospheric flux \citep[e.g.,][]{BasriBatalha1990ApJ...363..654B,Folha1999AA...352..517F}. 34,358 $r_\lambda$ values were collected from 5 publications for 9,588 sources. These are also summarised in Tab.~\ref{tab:acc}.

\subsection{Kinematic methods}\label{sec:kinematic}

Due to the short timescales involved in the star formation process, observed groups of stars born from the same cloud are still moving together in relation to the galactic centre, presenting kinematic properties common to the parent cloud. Hence, groups of young coeval stars can be distinguished from foreground and background sources based on their common kinematic properties. Kinematic methods - based on variations in the star's position in the sky - are thus widely used for selecting members of young coeval populations. There are two complementary techniques for studying stellar kinematics: the study of proper motions and the study of radial velocities. Today, reliable proper motion measurements are widely available due to \emph{Gaia}, including many studies that evaluate the membership of Gaia sources in the OSFC. Members of the OSFC identified by Gaia kinematic studies are included in our catalogue if they were part of a study focused on the OSFC or in star-forming regions \citep[e.g.][]{2019AJ....157..109K,2018AandA...620A.172Z}. Large-scale studies = typically based on machine learning, specialised in the identification of clustered populations all-sky \citep[e.g.][]{Hunt2021AA...646A.104H} were considered beyond our scope.  In addition, we also collated 60,294 radial velocity measurements for 13,721 sources from 33 publications. However, two caveats must be kept in mind. First, the inclusion of kinematic members of the OSFC will include its massive population (Sect.~\ref{sec:massivestars}). Second, radial velocities are susceptible to variations due to, for example, binarity, and these should ideally be examined in the context of an associated Julian date, which is often not reported in the studies included in our historical compilation. 

\subsection{Variability}\label{sec:var}

Since very early studies by \citet{Joy1945}, 
variability has been recognised as a defining characteristic of YSOs and has since been verified across the electromagnetic spectrum \citep[e.g.][]{Stelzer2007AA...468..463S,Cody_2014AJ....147...82C,Rebull2015AJ....150..175R,Venuti2015AA...581A..66V,Sousa2016AA...586A..47S,Roquette2020AA...640A.128R}. The OSFC has been historically pivotal in understanding the variability of YSOs \citep[e.g.,][]{Herbst1994,2001AJ....121.3160C}. Accordingly, we have included YSO candidates identified in the literature based on their variability traced back to the physical phenomenology behind the evolution of YSOs. 

\subsubsection{Variability amplitudes}\label{sec:var_amp}

Variability amplitudes are the most widely reported variability proxy in the YSO literature. 
Although amplitudes have been tabulated by a number of variability studies, we note that these values are highly inhomogeneous and may be challenging to interpret as an ensemble. As an alternative, rather than collating literature values, we further added value to the catalogue by using Gaia DR3 data to added the variability degree of our sources.

Although Gaia DR3 light curves are only available for $\sim25\%$ of sources in our catalogue, an alternative proxy for variability amplitude can be obtained from Gaia's DR3 mean flux uncertainties through the $A_{G\mathrm{proxy}}$ \citep[][see their eq. (2)]{Mowlavi2021AA...648A..44M}, which could be calculated for 24,623 sources in our catalogue using parameters from the Gaia DR3 mean photometry. 

\subsubsection{Stellar rotation observables}\label{sec:spin}

One of the most common variability processes in YSOs is the rotational modulation spots in the stellar surface, be it cool spots induced by magnetic activity, or hot spots induced by accretion. Measurements of variability periods are thus often associated with YSOs spin rates \citep[e.g.][]{2021ApJ...923..177S}. 

Due to this association with stellar rotation, we have grouped variability periods collated with $v\sin{i}$ measurements under a table focussed on stellar rotation observables (see Tab.~\ref{tab:Rot_data_types}).
42,133 such observables were available, for 16,774 sources from 35 publications. However, we disclaim that all variability periods collected from the literature are reported in Tab.~\ref{tab:Rot_data_types} under the variable name \texttt{Per} without discerning the possible physical origins for these periods. For example, eclipsing binaries and occultation by disc material outside the co-rotation radius can also explain periodic variability in YSOs.

\subsection{X-ray emission}\label{sec:xray}

High levels of X-ray activity are observed in YSOs at all stages of evolution from Class I to the ZAMS \citep{FeigelsonMontmerle1999ARAA..37..363F,Preibisch2005ApJS..160..401P}. Additionally, since the ratio between the bolomeric and X-ray luminosities for YSOs is expected to be $10^2-10^3$ larger than field stars with $M\sim 0.5-2~\mathrm{M}_\odot$ \citep[e.g.,][]{FeigelsonMontmerle1999ARAA..37..363F,Preibisch2005ApJS..160..390P,2005ApJS..160..353G}, X-ray observations have been established as a powerful tool for identifying YSOs. X-ray observations are especially powerful for identifying Class III, which typically lack most of the optical signatures of youth visible in Class I and II sources \citep[e.g.,][]{Walter1988AJ.....96..297W}. We identified 19 publications that provided X-ray observations for 4,162 likely YSOs in the OSFC. Whenever available, we collected X-ray luminosities, observed and/or corrected fluxes, hardness ratio, and hydrogen column density $\log{N_\mathrm{H}}$. More details on the data collected are provided in Tab.~\ref{app:tab:xray}.

\section{Multiplicity labels and possible duplicates}\label{sec:multiplicity}

We curated labels addressing the multiplicity of sources in our catalogue in four stages: through the historical compilation (1,539 sources labelled in Sect.~\ref{sec:bin:lit}), the use of all-sky binary catalogues from the literature (1,603 sources, Sec~\ref{sec:bin:largish}), the use of Gaia DR3 data (6,776 sources, Sect.~\ref{sec:bin:GDR3}) and derived using big data approaches (
6,912 sources, Sect.~\ref{sec:bin:bigdata}). As a result, 
18,930 \verb|bin_type| labels were assigned to 
12,155 sources, which can assume the following values:

\begin{itemize}
    \item \emph{S - Binary identified from Spectroscopy}: sources that were identified as spectroscopic binary (single or double-lined) or had observed variability in their radial velocity measurements associated with binarity. 
    \item \emph{E - Eclipsing Binary:} sources that were reported as such in studies focused on their light curves. 
    \item \emph{A - Astrometric Binary:} binaries identified based on their proper motion variations;
    \item \emph{B - Generic:} sources identified as a multiple system by previous studies being re-reported in a given study included in the compilation without enough accompanying information to be categorised elsewhere. 
    \item \emph{V - Visual pairs:} source identified as multiple from surveys with high angular resolution\footnote{We note that the term ``visual binary'' is often used interchangeably with ``visual pair''. The former term refers to physically associated double stars, which are observed close to each other in the sky as a result of their orbital movement. The latter refers to stars that appear close together in the sky, but are not necessarily physically associated.}.
    \item \emph{U - Unresolved pairs:} Unresolved close-pairs, typically indirectly identified using Gaia data (see Sect.~\ref{sec:bin:GDR3})
       \item \emph{Bl - blended sources}: sources likely blended with a nearby source\footnote{Blended sources are sometimes observed as visual pairs and sometimes as unresolved pairs.} (see Sect.~\ref{sec:bin:GDR3}).
    \item \emph{R - RUWE unresolved candidate} - Likely unresolved pairs identified using Gaia's RUWE;(see Sect.~\ref{sec:bin:ruwe})
    \item \emph{? - candidate:} sources reported as binary candidates because the detection criteria are close to the reported significance threshold.
\end{itemize}

\begin{figure}[htb]
    \centering
    \includegraphics[width=0.49\textwidth]{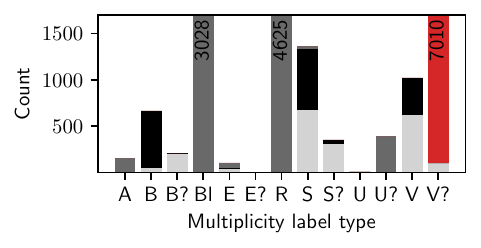}
    \caption{Incidence of different multiplicity labels included in the NEMESIS Catalogue of YSOs in the OSFC. Light grey bars show labels collected as part of the historical compilation (Sect.~\ref{sec:bin:lit}), black bars show labels collected from binary-focused catalogues (Sect.~\ref{sec:bin:largish}), dark gray bars show labels attributed using Gaia DR3 data (Sect.~\ref{sec:bin:GDR3}), and red bars show labels attributed using Big Data approaches (Sect.~\ref{sec:bin:bigdata}).} \label{fig:bin_lit}
\end{figure}

Figure~\ref{fig:bin_lit} presents the relative numbers of sources in each category. Some sources have been assigned to multiple categories. For example, 28\% of the sources reported as eclipsing binaries in the literature were also reported as binaries from spectroscopic investigations. We note that we have not attempted any distinction between binary or triple systems. 

\subsection{Multiplicity labels in from the historical compilation}\label{sec:bin:lit}

The historical data compilation described in
 Sect.~\ref{sec:data-historical} yielded the identification of 2,006 multiplicity labels for 1,539 sources (see also Fig.~\ref{fig:bin_lit}) collected from 93 publications. 

\subsection{Multiplicity labels in from all-sky catalogues}\label{sec:bin:largish}

We further searched the literature for large catalogues of binary stars that included the OSFC field and matched these to our source list. This helped us to attribute multiplicity labels to 1,735 sources, 1,102 of which were unlabelled in the historical compilation. This included:

\begin{itemize}
    \item (V?) 397 spatially resolved binaries with separations up to $\sim$ 200 AU identified with Gaia eDR3 data \citep{El-Badry2021MNRAS.506.2269E} with low chance alignment ;
    \item (E) 11 eclipsing binaries identified using the first 2 years of TESS observations \citep{Prsa2022ApJS..258...16P};
    \item (S) 31 double-lined spectroscopic binaries identified by \citet{Kovalev2024MNRAS.527..521K} using $v\sin{i}$ values from spectral fits to LAMOST-MRS spectra;
    \item (S?) 460 double-lined spectroscopic binaries identified by \citet{Zheng2023ApJS..266...18Z} using a deep-learning approach to analyse LAMOST-MRS spectra;
    \item (S?) 36 spectroscopic binary candidates or RV-variables identified by \citet{Qian2019RAA....19...64Q} due to their large radial velocity variations in their LAMOST-LRS spectra;
    \item (S) 19 double-lined spectroscopic binaries identified by \citep{Traven2020AA...638A.145T} with spectra from the GALAH survey;
    \item (S?) 2 spectroscopic binary candidates identified by \citet{Birko2019AJ....158..155B} due to their radial velocity variability in the RAVE survey;
    \item (S?) 7 spectroscopic binary identified by \citet{Dannis2019AN....340..386J} from a sample combining Gaia DR2 RV and RAVE spectra;
    \item (B?) 10 ellipsoidal variable candidates selected from their TESS light-curve variability \citep{Green2023MNRAS.522...29G};
    \item (B) 615 sources from the Identification List of Binaries Catalogue, which provided a survey of surveys of literature focusing on binary stars prior to 2015 \citep{Malkov2016BaltA..25...49M};
    \item (S) 148 sources from GALAH DR4 \citep{Buder2024arXiv240919858B} with a secondary component detected and for which primary and radial velocity could be measured.  
\end{itemize}

\subsection{Big data identification of visual-pair candidates}\label{sec:bin:bigdata}

In Sect.~\ref{sec:VizierSED}, we have employed the VizieR-SED tool to retrieve the photometric data available on CDS around the position of each source in our catalogue. In this section, we further employ these data to gauge sources' neighbourhoods. 

\begin{figure}
    \centering
    \includegraphics[width=0.9\linewidth]{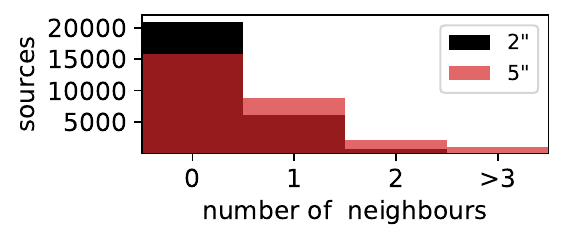}
    \caption{Distribution of the number of neighbours around sources in the NEMESIS Catalogue of YSOs in the OSFC. Results for a searching radius of 2" are shown in red, and 5" in black. \label{fig:n_neighbours}}
\end{figure}

Starting from the raw VizieR-SED output for a 5" searching radius, rather than relying again on the procedure described in Sect.~\ref{sec:SED} we carried a simpler preprocessing. We kept only data in the wavelength range 0.01-5$\mu m$, excluding multi-epoch surveys (e.g., TESS) and surveys with source detection and extraction at different apertures (e.g., SDSS). We then grouped the remaining data by unique CDS table name and photometric band and counted how many unique pairs of RA, Dec (down to a $10^{-7}$ precision) existed. This process was repeated for each pair of CDS table name and photometric band, and the maximum number of unique sources inside the searching radius was stored as representative of the neighbourhood of that source. The same procedure was repeated using a 2" searching radius as well. The distribution of the number of neighbours for sources in our catalogue is shown in Fig.~\ref{fig:n_neighbours}. 

This procedure allowed us to leverage the large amount of photometric data available within CDS to investigate the neighbourhood of sources in our catalogue, including data observed by instrumental setups yielding a wide variety of pixel scales and PSF sizes, while being agnostic to the selection window introduced by the criteria for inclusion in this compilation discussed in Sect.~\ref{sec:historical-dtypes}. 
We found that 6,912 sources in our catalogue have at least one neighbour within 2" from the sources' adopted coordinates, and 12,060 have neighbours within 5". 
 Sources with at least one neighbour within 2" were labelled as visual pair candidates (\texttt{V?}) with the support that 71\% of these sources were also identified as such by other multiplicity labels considered. 

\subsection{Multiplicity labels in from  Gaia DR3}\label{sec:bin:GDR3}

The numerous avenues for investigating sources' multiplicity using Gaia DR3 data merit this dedicated section.

\subsubsection{Non-single source tables in Gaia DR3}
We identified 223 sources with a counterpart in at least one of a series of tables focussed on non-single stars released as part of Gaia DR3 \citep{Binary2023AA...674A..34G}, which are composed of sources identified as astrometric \citep[150 sources from $nss\_acceleration\_astro$;][]{Halbwachs2023AA...674A...9H}, spectroscopic \citep[18 sources from $nss\_non\_linear\_spectro$;][]{Gosset2024arXiv241014372G} or eclipsing binaries \citep[57 sources from $vari\_eclipsing\_binary$;][]{Mowlavi2023AA...674A..16M}, and with orbital models for all sources compatible with a two-body solution \citep[10 sources from $nss\_two\_body\_orbit$;][]{Holl_binary_2023AA...674A..10H}. 

\subsubsection{Gaia's scan-angle effects:}

Gaia's readout window has a different size depending on the magnitude of the source and photometric band. For example, the typical window for a faint source in $G$ is $\sim 0.35"\times 2.1"$, and $\sim 3.5"\times2.1"$ for $G_{BP}$ and $G_{RP}$ \citep[e.g.,][]{Riello2021_GDR3,Holl2023AA...674A..25H}. In addition to this ``rectangular'' window, to achieve full-sky coverage, Gaia has a non-trivial scanning law, including a time-varying scan-angle. While for isolated point sources, these scan-angle variations are irrelevant, for close pairs, they are determinant factors between the source being resolved or blended. For example, pairs with separations $\sim 1"$ will always be blended BP and RP while sources with separations $\sim 1 - 2"$ will occasionally be resolved when observed close to along-scan angles and can therefore be identified using Gaia's quality control flags. Meanwhile, pairs with separations $\sim 0.6 - 2"$ are sometimes resolved with a six-parameter solution for the G-band \citep[see also their Fig. 6][]{Lindegren2021a_GDR3}.

\paragraph{Scan-angle-dependent signals:}
Gaia's time-varying scan-angle is hypothesised to induce scan-angle-dependent signals when observing unresolved binaries or extended sources. We followed \citep{Holl2023AA...674A..25H} procedure to identify 191 sources likely susceptible to scan-angle biases compatible with unresolved pair numerical models (separations $\lesssim 1000$ mas), which we labelled as \emph{U?} (candidate unresolved pairs). 

\paragraph{Scan-angle-dependent spurious variability:}
Gaia's time-varying scan angle can also result in specific spurious periodic signals biasing Gaia's epoch data. \citet{Holl2023AA...674A..25H} and \citet{Distefano2023AA...674A..20D} attempted to identify such spurious periodicity by quantifying the correlations between Gaia's epoch photometry and: (i) the image direction
parameter determination goodness of fit ($spearman\_corr\_ipd\_g\_fov$), and (ii) the epoch-corrected excess factor ($spearman\_corr\_exf\_g\_fov$); with these parameters released as part of DR3's table \verb|vari_spurious_signals|. We identified 198 candidate unresolved sources (\emph{U?}) by selecting sources with a correlation greater than 0.8 in both of these parameters. 

\paragraph{Blended sources:}

Due to Gaia's several instruments and its different window sizes for the $G$, and the $G_{BP}$ and $G_{RP}$ bands, sources resolved in the first may be blended in the others. Gaia DR3 data processing provides flags that attempt to quantify the number of transits in which a source was blended with a nearby source. \citet{Riello2021_GDR3} employs these flags to define a metric, $\beta$ \footnote{Following \citet{Riello2021_GDR3}, $$\beta=\frac{{phot\_bp\_n\_blended\_transits}+{phot\_rp\_n\_blended\_transits}}{{phot\_bp\_n\_obs}+{phot\_rp\_n\_obs}}$$, where $phot\_?\_n\_blended\_transits$ is the number of transits in the $?$ band (BP or RP) contributing to Gaia's mean photometry in that band, which were flagged to be blends of more than one source inside the observing window; and $phot\_?\_n\_obs$ is the total number of observations contributing that band's mean photometry.}, to gauge the blending fraction of source. We used this metric to flag as \texttt{Bl} (blended) 3,028 sources that had more than 20\% of their Gaia DR3 transits flagged as such ($\beta\geq0.2$).

\subsubsection{Gaia's Renormalised Unit Weight Error}\label{sec:bin:ruwe}

When observed by astrometric surveys like Gaia, stars with an unresolved or faint companion will show a different motion in their centre of light and centre of mass, which yields poor results when attempting to fit single-source astrometric models. In Gaia, such sources can be identified in terms of their Renormalised Unit Weight Error (RUWE), which is expected to be close to one for single sources with a well-defined centre of light and uniform motion \citep{Lindegren2018RenormalisingTA}. Since DR3, the threshold of $ruwe=1.4$ \citep{Lindegren2018RenormalisingTA} widely used with DR2 data to distinguish between single and multiple sources has been the subject of debate. For example, studies focused on eclipsing binaries have found a strong correlation between RUWE values and the separations of the unresolved binaries down to $ruwe=1$ \citep[see their fig.~3;][]{Stassun2021ApJ...907L..33S}. 
As part of the \emph{GaiaUnlimited}\footnote{\url{https://gaia-unlimited.org/}}, \citet{2024AA...688A...1C} provide a model to estimate suitable thresholds for binary selection with RUWE as a function of Gaia's selection function variations as a function of sky position. This allowed us to choose appropriate RUWE values at the position of each source in our catalogue. We found an average threshold of \texttt{ruwe} $1.216\pm0.038$ for sources in our catalogue. \emph{GaiaUnlimited} also provides tools to estimate the probability of erroneous selection of a single star with RUWE as a function of sky position, with a $5\pm3\%$ probability in the OSFC field. With 24,635 sources in our catalogue having at least one Gaia DR3 counterpart within 2", and 23,716 sources having a \texttt{ruwe} measurement in DR3, 4,625 sources were flagged \emph{R} based on our RUWE thresholds and should be interpreted as probable unresolved multiple systems.

\subsubsection{Gaia's RUWE in the context of a large sample of YSOs}

As an illustrative application of our catalogue, we further contribute to probing previous suggestions that Gaia's RUWE may be inflated in YSOs. As discussed in Sect.~\ref{sec:bin:ruwe}, Gaia's astrometric solution assumes that sources are point-like with a well-defined centre of light and a single-star movement. Along with multiple systems, extended sources also deviate from these assumptions due to their non-point-like nature and the underlying larger uncertainty in resolving their centre of light. In YSOs, the diffuse emission of the envelope in Class 0/I sources, or even the radiation re-emitted by a thick disc in Class II sources could be the culprit of an uncertain centre of light.

\paragraph{Does the presence of a disc inflate Gaia's RUWE?}\label{sec:ruwe_disc}

The inflation of RUWE values due to the presence of a circumstellar disc has previously been suggested by \citet{Fitton2022RNAAS...6...18F}, based on a modest sample of 122 single stars selected from adaptative optics surveys for which they find an excess of disc-bearing sources in the RUWE range 1--2.5. 
Here, we contribute to probing this evidence by examining a much larger sample of YSOs. For this end, we selected sources from our catalogue with both $\alpha_{IR}$ derived in Sect.~\ref{sec:alphaindex}, and Gaia's DR3 RUWE. 

We defined two subsamples based on $\alpha_{IR}$ (Sect.~\ref{sec:alphaindex}), namely sources with a thick disc ($\alpha_{IR}\geq-1.6$) and discless sources ($\alpha_{IR}\leq-2.5$). To maximise the reliability of sample selections using the $\alpha_{IR}$ indices, we considered only sources with at least one datapoint available for $\lambda\geq5\mu$m.
We note that the requirement for Gaia data will inevitably remove the least evolved and most embedded Class I/O sources from our catalogue, as they are too faint and beyond Gaia's detection capacity. 
To maximise sample purity, we also removed all sources flagged as possible contaminants in Sect.~\ref{sec:contamination}.
Next, we also removed all sources flagged as multiple in Sect.~\ref{sec:multiplicity}, except for cases where the only multiplicity flag was the RUWE-based one, \texttt{R} (Sect.~\ref{sec:bin:ruwe}). This is required to preserve sources in the RUWE range that \citet{Fitton2022RNAAS...6...18F} claim to be populated by disc-bearing sources. Although 51\% of the \texttt{R} labelled sample is still removed, as they were also identified as multiple systems by other methods, a modest fraction of $\sim10\%$ sources with RUWE$>2.5$ and probably real unresolved systems are left in the sample. 
To further minimise the influence of remaining binaries in our results, we focused our analysis on the RUWE$<$2.5 range, which has been suggested to be the relevant range for disc-bearing sources \citep[see][]{Fitton2022RNAAS...6...18F}. Our final sample was composed of 4\,364  discless sources and 1\,489 thick disc sources.

We employed a series of non-parametric statistical tests (Kolmogorov-Smirnov, Mann-Whitney U, and Anderson-Darling) to examine the differences between RUWE distributions of disc-bearing and discless sources, finding no support for significant differences between the two distributions. Alternatively, we have also explored other criteria for sample selection (e.g., the use of $W1-W3$ colours for thick disc selection) and other statistical approaches, such as examining disc fraction as a function of RUWE, but found no statistically significant indication of RUWE inflation due to the presence of thick discs around less evolved YSOs in our sample.

\paragraph{Does the variability of YSOs inflate RUWE?}\label{sec:ruwe_var}

Beyond extended sources, photometric variability has also been proposed as a contributor to RUWE. For example,  \citet{Belokurov2020MNRAS.496.1922B} previously uncovered a trend of inflated RUWE as a function of variability amplitude for RRLyrae and Cepheids observed with Gaia DR2. As further discussed by these authors, the inflation of RUWE by variability can be traced back to RUWE's definition \citep[][eq. (4)]{Lindegren2018RenormalisingTA}. As its name says, RUWE is a normalised version of Gaia's 5-parameter astrometric solution's unit error, where a normalisation coefficient is derived from one of the modes UWE's as a function of magnitude and colour. In variable stars where sharp changes of magnitude and colour take place, this normalisation may be incorrect, yielding spurious inflation of RUWE. Since variability is a prevalent characteristic of YSOs (Sect.~\ref{sec:var}), here we also address the influence of variability on Gaia's DR3 RUWE.

\begin{figure*}
    \centering
    \includegraphics[width=0.9\linewidth]{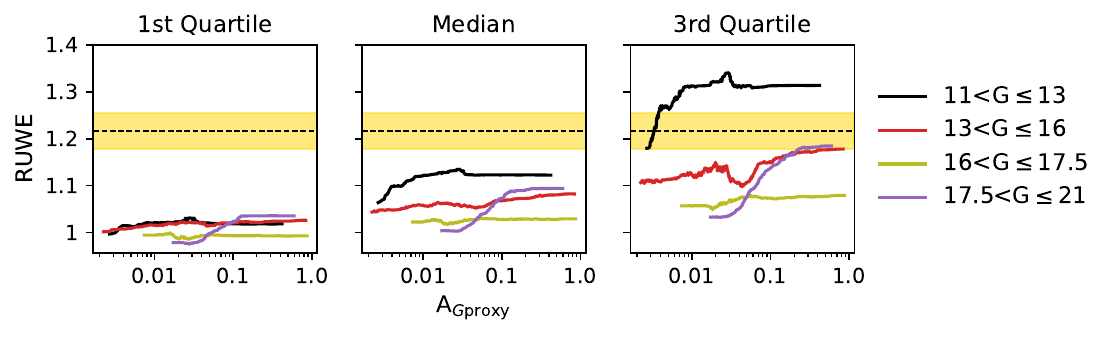}
    \caption{RUWE as a function of variability amplitude ($A_\mathrm{G proxy}$) in Gaia DR3. A sample of YSOs (without known binaries and restricted to RUWE$<2.5$)  is divided into 4 magnitude bins, with different rolling quartiles shown in the left (1st quartile), middle (median) and right (3rd quartile) panels.
     In all panels, the yellow dashed region and black-dashed line show the threshold for selection of unresolved pairs with RUWE adopted in this study.
    }
    \label{fig:ruwe_variability}
\end{figure*}

We used Gaia DR3 $A_{G\mathrm{proxy}}$ calculated in Sect.~\ref{sec:var_amp} (available for $\sim 88\%$ of the sources in our catalogue) to address the evidence for variability influence on RUWE \citep[e.g.][]{Belokurov2020MNRAS.496.1922B,Barber2023ApJ...953..127B}. As in the previous section, we minimised the incidence of binaries in the sample by removing all sources labelled as such, except for sources with only a \texttt{R} flag, and examining RUWE values up to 2.5. We also removed all sources flagged as possible contaminants in Sect.~\ref{sec:contamination}, which left us with 10,326 sources. Next, we bin the data at magnitude ranges $11\leq G<13$ (848 sources), $13\leq G<16$ (3,571), $16\leq G<17$ (3,850), and $17\leq G<21$: (2,057) with the first two bins following Gaia's window classes, and the remaining aimed at minimising the effects of the signal-to-noise variations with magnitude. For each magnitude bin, we estimate the first, second, and third rolling quartiles for the distributions of RUWE as functions of $A_{G\mathrm{proxy}}$. The results are shown in Fig.~\ref{fig:ruwe_variability}, where variations of $\approx0.03-0.18$ in RUWE are verified as a function of amplitude.  

\subsection{Possible Duplicates}\label{sec:duplicate}

After extensively evaluating the multiplicity of sources in our catalogue, we are now equipped to address the incidence of duplicated sources in our data compilation process. An internal match with sources' sky positions and a 2" radius reveals that 1,864 sources are possibly duplicated. We verified that 1,687 of these had a multiplicity flag derived in this section or had their possible duplication flagged as multiple. From the remaining, 119 sources could be traced back to the publications introducing them to our dataset in Sect.~\ref{sec:data-historical}, which were typically surveys including higher-resolution images, and we could confirm that multiple sources were indeed reported by their original publication, 4 had support for being independent sources from a match with Simbad, and 20 had a neighbourhood flag (Sect.~\ref{sec:bin:bigdata}) indicative of at least one neighbour within a 5" radius. Finally, only 34 sources could not be explained and are likely duplicates due to some coordinate mismatch in our historical data compilation. These sources were flagged accordingly. 

\section{Catalogue purity and contamination assessment}\label{sec:discussion}

\subsection{Comparison with Simbad}\label{sec:comparison-simbad}

We used Simbad's TAP service (on 13th December 2024) to retrieve data for 165\,831 sources indexed by Simbad and located in the same field as our survey. We used a 5" searching radius to search for possible counterparts at the position of each source in our catalogue. This relatively large searching radius was chosen to account for possible offsets between coordinates adopted in the historical compilation and the ones adopted by CDS\footnote{It was noted during the data integration process (Sect.~\ref{sec:data-integration}) that infrared source coordinates (e.g., from 2MASS) sometimes show offsets with optical coordinates (e.g., from Gaia) that cannot be explained by the different epochs of observations. This was particularly relevant for YSOs with large envelope contributions or part of multiple systems. As a cautionary example, we point to the star V 1118 Ori (NEMESIS \texttt{Internal\_ID} 5465), a widely studied Class II YSO reported in our database with its 2MASS coordinate (\emph{2MASS J05344474-0533421}). V 1118 Ori is indexed by Simbad with its ICSR J2000 coordinates from Gaia DR2, which are offset by 3.67" from its 2MASS coordinates.}. We found 33\,846 sources indexed by Simbad with a possible counterpart in our catalogue. 25\,902 sources in our catalogue have at least one Simbad counterpart within the searching radius. All possible counterparts were recorded and are reported in our catalogue, regardless of their Simbad object type label (Tab.~\ref{app:tab:main}), including 3\,167  sources with multiple possible counterparts. We leave it to the users of the catalogue to rank possible counterparts according to their use case.

\subsubsection{Young sources known as such by Simbad} 

We followed \citet{Simbad_guide} guidelines on object types to identify 19\,625
sources in the OSFC field with \emph{object\_type} labels indicative of T Tauri Stars (\texttt{TT?} or  \texttt{TT*}), Young Stellar Object (\texttt{Y*?}, \texttt{Y*O}), Ae/Be Herbig stars (\texttt{Ae*} or \texttt{Ae?}), or Orion Variables (\texttt{Or*}, which variable stars with irregular variability associated with eruption phenomena in young stars). We note that this list does not include objects that are arguably related to star-forming regions, such as dense cores (\texttt{cor}) and Herbig Haro objects (\texttt{HH}). Although some of these sources are indirectly included in our catalogue, as discussed in Sect.~\ref{sec:data-historical}, they were considered outside the main scope of our compilation.
Here, we call the first group of Simbad's object type labels \emph{"YSO labels"}, i.e. labels that directly imply the source is a YSO. We call the second group \emph{"star-formation labels"}, i.e. labels that imply that the source is part of a star-forming region. Considering both sets of labels, 18\,029 sources in our catalogue were already known by Simbad as young. Among the 7\,873 sources in our catalogue with a close Sibmad counterpart but not in the two categories of labels considered, 87\% were generically labelled as star (\texttt{*}).

\subsubsection{YSOs known by Simbad but missed by our compilation}

Among the YSOs known by Simbad, 2\,112 were not included in our catalogue. 280 of these are located in Mon R2, which was excluded from our survey. We have retrieved the bibliographic references associated with the remaining sources and verified that 1,005 of them originate from studies with titles and abstracts focused on clumps and cores, which were considered out of our scope in Sect.~\ref{sec:data-historical}. 251 sources could be traced back to CDS tables included in our compilation but were not deemed YSO candidates by the criteria in Sect.~\ref{sec:historical-dtypes}. The remaining sources were associated with one of the following cases. They came from all-sky variability surveys including ML-generated YSO variability flags \citep[e.g.][]{2017ARep...61...80S}; they came from all-sky cluster membership studies with Gaia data, whereas we have only included such surveys in our compilation when they were specifically tailored for studying the OSFC or star-forming regions \citep[e.g.][]{2018AandA...620A.172Z}; they came from studies focused on the OSFC but did not include tables on CDS and had too few sources for their tables to make it to the list of tables digitalised by us in Sect.~\ref{sec:histDataRetrival}; They came from studies focused on YSOs, but with bibliographic entries absent from the searching terms utilised in Sect.~\ref{sec:NASA/ADS}. Finally, we identified one bibliographic entry that adds a few dozen YSO candidates \citep{2002ApJ...573..366M}, which was discarded in Sect.~\ref{sec:histDataRetrival} due to its focus on massive stars.

\subsubsection{Sources unknown to Simbad} 

Our catalogue included 1\,977 sources without a counterpart in Simbad. It is interesting to note that although these sources were unknown to Simbad, 90\% of them had data collected with the VizieR-SED tool within a search radius of 2". 95\% of these sources had at least 4 data points in their compiled SEDs, but some had as many as 50 data points. Beyond their SEDs, 1\,623 sources unknown to Simbad had measurements of at least one of the relevant observables discussed in Sect.~\ref{sec:historical-dtypes}, 
with 89\% having $\alpha_{IR}$ in Sect.~\ref{sec:alphaindex} comprising 477, 847, 140, 55, and 237 were classified as discless Class III, thin-disc Class III, Class II, Flat-Spectrum and Class 0/I respectively. The 125 sources unknown to Simbad that did not include any relevant observables related to youth other than photometry could be traced back to publications included in our compilation that selected YSOs using a CMD, with 42 of these sources from \citet{2009ApJS..183..261D}, 70 from \citet{2019MNRAS.486.1718S} and  42 from \citet{2014AandA...564A..29B}.

\subsection{Non-YSOs and catalogue purity}\label{sec:contamination}

\begin{figure*}
    \centering
    \includegraphics[width=0.9\linewidth]{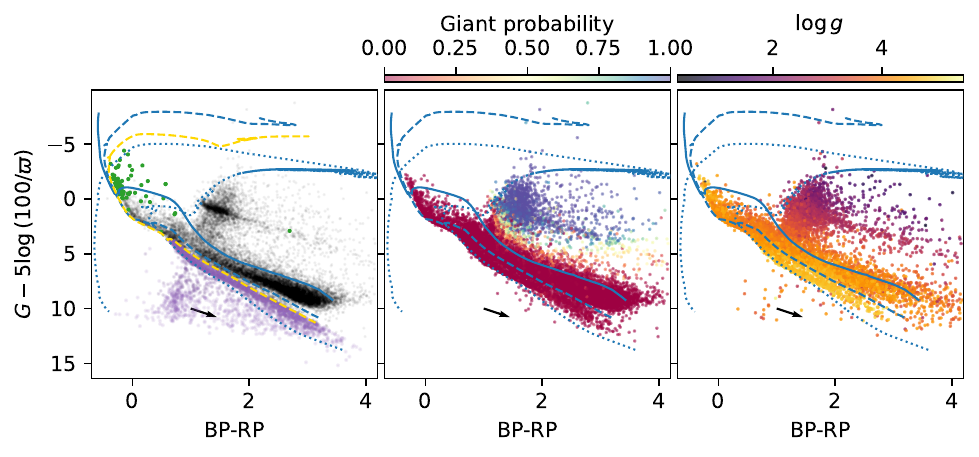}
    \caption{Gaia DR3 CMD for sources in the NEMESIS Catalogue of YSOs in the OSFC (shown as black dots). Blue lines show isochrones for 1 Myr (continuous line), 10 Myr (dashed line) and 1 Gyr (dotted line). The reddening direction is indicated by the black arrow, which has an $A_V=1$ mag length.  \emph{Left:}  Massive stars (Sect.~\ref{sec:massivestars}) are shown as green dots. Likely MS and post-MS contaminants in our catalogue are shown as purple dots below a 20 Myr isochrone (yellow-dashed). \emph{Middle:} Sources are coloured by their Giant probability derived with a Random Forest binary classifier in App.~\ref{app:galaxy}. \emph{Right:} Sources with $\log{g}$ collated from spectroscopic surveys (Sect.~\ref{sec:spec-surveys}) are shown with coloured by their $\log{g}$ measurement (Sect.~\ref{sec:giants}).}
    \label{fig:GDR3_CMD_Giants}
\end{figure*}

In Sect.~\ref{sec:historical-dtypes}, we employed a positive evidence rule to compose our catalogue, where a single peer-reviewed study reporting a source as a YSO candidate was enough to grant its inclusion here. We have thus prioritised the completeness of our YSO census, rather than its purity. Consequently, YSO purity must be addressed depending on the intended use of the catalogue.
Although a formal evaluation of the reliability of youth indicators of each individual source in the catalogue is beyond the scope of this paper and the subject of a follow-up study, here we discuss a series of assumptions to assess our catalogue's degree of contamination by different types of sources. To support this discussion, Fig.~\ref{fig:GDR3_CMD_Giants} shows a Gaia DR3 CMD for 23,716 sources in our sample with Gaia photometry and parallaxes. As a reference, we used solar-metallicity MESA isochrones \citep{Dotter2016ApJS..222....8D}.

\subsubsection{Massive stars}\label{sec:massivestars}

We call \emph{massive stars} high-mass stars that are past the PMS phase. From a combination of the adopted MESA stellar evolution models with \citet{PecautMamajek2013} $T_\mathrm{eff}$-Spectral Type scales, and considering sources at 1 Myrs, this is equivalent to sources with spectral type earlier than B3. Based on the spectral types collated in Sect.~\ref{sec:SpT}, 80 sources in our catalogue met this criterion (see Fig.~\ref{fig:GDR3_CMD_Giants} left panel), yielding a $\sim0.3\%$ incidence of massive stars. 

\subsubsection{Contamination by Giants}\label{sec:giants}

Because background reddened giants are a known major pollutant to YSO identification studies, we wish to further add value to our catalogue by ranking possible giant contaminants. In Sect.~\ref{sec:gravity} we collected 57\,649 $log{g}$ measurements for 12\,313 sources in our catalogue. Although giant contamination in our catalogue was already hinted at in Fig.~\ref{fig:logg}, the direct use of these measurements for the purpose of identifying giants is hampered by the wide spread in $log{g}$ derivations from different surveys. Thus, we used $log{g}$ uncertainties as a statistical weight when combining multiple literature values and restricted our analysis to 6\,895 sources with either a 3$\sigma$ dispersion or $log{g}$ uncertainty (if only one measurement is available) below 0.1 dex. Fig.~\ref{fig:GDR3_CMD_Giants} (right panel) shows the distribution of $log{g}$ for this sample, which yields the identification of 537 likely giants with $log{g}\leq2.5$ (this threshold was chosen based on distributions of $log{g}$ values in the MESA evolutionary models with the aim of minimising overlap in $log{g}$ distributions of giants and YSO).

To expand the giant contamination investigation to a larger portion of our catalogue, in App.~\ref{app:galaxy} we employed giants identified based on $\log{g}$ estimation from large spectroscopic surveys to implement a Random-Forest binary classifier trained on features from Gaia DR3 to estimate the probability that sources in our catalogue have properties comparable to a known sample of Giant stars. The results of this giant classifier for our sample are shown in the middle panel of Fig.~\ref{fig:GDR3_CMD_Giants}. We were able to derive such probabilities for 20\,546 sources in our catalogue (reported in Tab.~\ref{app:tab:main}), and 2\,614 sources with high probabilities of being a Giant. We thus estimate a giant contamination level of $\sim 13\%$ in our catalogue.
We note that among high giant probability sources, 515 sources were in the sample detected from their logg. Altogether, sources with logg indicative of giants and sources identified by our classifier correspond to 9.5\% of our catalogue, suggesting a small residual incidence of unlabelled giants.

\subsubsection{Extragalactic contamination}\label{sec:galaxy}

As part of another study within the framework of NEMESIS \citep{NEMESIS_David}, we visually identified 11 sources with galaxy morphology in photometric images. One of these sources, 2MASS J05401945-0713593 (former \texttt{Internal\_ID}=4281) had a counterpart in the NASA/IPAC Extragalactic Database and was removed from our compilation as soon as identified. Another source, 9334 (HOPS 292), has been pointed out as a possible Galaxy in the YSO literature. The other 9 (4330, 4331, 4547, 4608, 4609, 7319, 7471, 7492, 7533) are new identifications of Galaxies, with 8 of these previously reported as YSO by Simbad. 

\begin{figure}
\centering
    \includegraphics[width=0.9\linewidth]{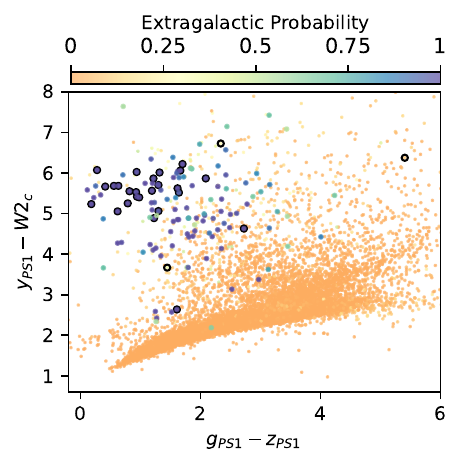}
    \caption{Colour-colour diagram for the $y_\mathrm{PS1}-W2_\mathrm{cat}$ vs. $g_\mathrm{PS1}-z_\mathrm{PS1}$, which are two of the features used in the Extragalactic classifier discussed in App.~\ref{app:galaxy}. The colour bar reflects the derived probability of a source being likely extragalactic. Black circles show extragalactic sources reported in the literature.}
    \label{fig:GalaxyContamination}
\end{figure}

In App.~\ref{sec:GalLit}, we collated a sample of known extragalactic sources in the field of the OSFC, including resolved and unresolved Galaxies, AGNs and Quasi-stellar objects or Quasars (QSOs). 
Among this literature-based extragalactic sample, 34 sources had a close counterpart in our catalogue. The full labelled sample was used in the implementation of a Random Forest binary classifier trained to identify extragalactic sources based on photometric colours from the Gaia DR3, PanSTARRs DR2, and CatWISE surveys. The trained classifier was used to 
 evaluate the probability of 15\,855 sources in our catalogue having properties comparable to the known sample of extragalactic sources. This allowed us to identify 140 likely extragalactic sources in our catalogue, 115 of those previously unlabelled extragalactic, yielding an estimated incidence of $\sim 0.9\%$ extragalactic contaminant in our sample. Fig.~\ref{fig:GalaxyContamination} shows one of the parameter spaces defined by the features used in our classifier. Altogether, we identified 159 probable extragalactic sources in our sample ($\sim0.6\%)$, suggesting a small residual incidence of extragalactic contaminants.

We note that the misclassification of extragalactic sources and YSOs happens both ways. 
During the collation of a labelled sample for our extragalactic classifier in App.~\ref{app:galaxy}, we refrained from including the Gaia DR3 QSO candidates in the labelled sample due to evidence of a relevant incidence of YSOs in this sample. Following Gaia DR3's documentation \citep{GDR3_Extragalactic_2023AA...674A..41G} for selecting a high-purity QSO sample towards the OSFC, we found 81 QSO candidates in common with our YSO catalogue. Further examination of these sources on light of the diverse attributes included in our database revealed that a relevant fraction of such sources had compelling evidence for bona fide YSO classification in our catalogue, with 35 of these having spectroscopic constraints indicative of youth. Out of the 46 remaining, 30 were identified as high probability extragalactic contaminants in App.~\ref{app:galaxy}, but 16 are unaccounted. While we opt not to include this sample in our contaminant sample, users of the catalogue who wish for a full removal of possible extragalactic sources are recommended to use the \texttt{extragalactic\_label} and \texttt{strat\_label} (Tab.~\ref{app:tab:main}) to identify and further inspect such sources.

\subsubsection{Further contamination}\label{sec:ms_contamination}

As also suggested by our comparison with Simbad, Fig.~\ref{fig:GDR3_CMD_Giants} also reveals contamination by low-mass MS stars. We isolated this contamination based on a 20 Myr MESA isochrone, finding 2\,845 sources below the isochrone. Based on the amount of Gaia DR3 good-quality data available, this suggests a $\sim 12.5$~\% contamination incidence.

\section{Summary of Results}

We have assembled the largest historical compilation of YSOs in the Orion Star Formation Complex to date, including 27\,879 previously reported YSO candidates. The list of YSO candidates reported here included:

\begin{enumerate}
    \item Young sources identified through HRD or CMD analysis or on the bolometric luminosity-temperature relations of less evolved YSOs, including $T_\mathrm{eff}$ or $T_\mathrm{bol}$ measurements for 63\% sources and Spectral Types of 41\%.
    \item Sources identified in photometric and spectroscopic studies focused on infrared emission of circumstellar material, including IR classification for 35\% of sources.  
    \item 78\% of sources included spectroscopic features associated with the star-disc-interaction and accretion processes in YSOs; 
    \item Youth-related spectral absorption lines such as Lithium and certain gravity-sensitive spectral lines for 22\% and 48\% of sources; 
    \item Sources identified by kinematic studies focused on the OSFC populations including RV measurements collated for 49\% of sources.
    \item Variability amplitudes for 88\%, and variability periods collated for 36\% of sources.
    \item X-ray emission traced back to YSOs, with X-ray luminosities and/or fluxes collated for 14\% of sources. 
\end{enumerate}

Beyond data collated from 217 previous publications, the catalogue presented here was also complemented with panchromatic data from both photometric and spectroscopic large-scale surveys. We further added value to our catalogue by: 

\begin{enumerate}[resume]
    \item compiling SEDs for all sources in the catalogue, with 96\% of sources with at least 5 data points in their SEDs;
    \item homogeneously deriving $\alpha_{IR}$ indices and IR classes for 93\% of sources; 
    \item extensively evaluating the multiplicity, resulting in 43\% sources in the catalogue labelled as likely multiple sources (binary candidates or visual pairs). 
\end{enumerate}

Due to the positive evidence approach adopted for our data collection, our catalogue prioritises completeness over purity. We estimate that {\bf $\sim73\%$} of the sources in our catalogue are reliable YSOs, with 0.3\% of the catalogue is composed of massive stars, 13\% are likely giants contaminants, $\sim1\%$ are likely extragalactic contaminants, and 12.5\% are likely MS contaminants. Finally, we have estimated that the incidence of duplicates in this catalogue is below the 1\% level. 

As an illustrative application of the catalogue, we add to the evidence that Gaia's RUWE parameter - commonly used for identifying unresolved binaries - may be inflated in YSOs due to their large variability. 

The present catalogue is already in use by our collaborators for upcoming publications related to the NEMESIS project. \citep{NEMESIS_David} employed the catalogue to select a list of YSOs to investigate the morphology of YSOs using self-organising maps; \citep{Marton2024eas..conf.1922M, NEMESIS_YSO_Gabor} used the catalogue as a training set for the identification of YSOs all-sky using deep learning techniques; and \citep{NEMESIS_Chloe} used the YSO observables collated here to validate the results concerning the variability of YSOs observed by Gaia DR3. \citep{NEMESIS_Ilknur} presented further YSO parameters by fitting YSOs' SEDs with disc models. An ongoing study will further add value to the present catalogue through the ranking of bona fide YSOs (Roquette et al. in preparation).

\begin{acknowledgements}
We acknowledge funding from the European Union’s Horizon 2020 research and innovation program (grant agreement No.101004141, NEMESIS).
G.M. acknowledges support from the János Bolyai Research Scholarship of the Hungarian Academy of Sciences. This research was also supported by the International Space Science Institute (ISSI) in Bern, through ISSI International Team project 521 selected in 2021, Revisiting Star Formation in the Era of Big Data (\url{https://teams.issibern.ch/starformation/})
We thank Berry Holl for helpful discussions on the use of Gaia DR3 data, Lynne Hillenbrand for exchanges about large-scale collection of YSOs, and Sotiria Fotopoulou and Javier Acevedo Barroso for illuminating discussion on the identification of extragalactic sources. 
This work made use of Astropy \citep{astropy:2013, astropy:2018, astropy:2022}, Scikit-learn: Machine Learning in Python \citep{scikit-learn}, Pandas \citep{reback2020pandas}, Numpy \citep{harris2020array}, Natural Language Toolkit for Python \citep{bird2009natural}, and Matplotlib \citep{Hunter:2007}.

\end{acknowledgements}

%
   \bibliographystyle{aa} 
   \bibliography{ref} 
\begin{appendix} 

\section{ALLWISE Cleaning}\label{app:allwise}

The WISE mission \citep{Wright2010_WISE} was designed to target infrared-bright galaxies, brown dwarfs, and near-Earth asteroids. Hence, besides the convenience of its mid-infrared full-sky coverage, WISE observations of YSOs in embedded environments are often susceptible to spurious detections, especially towards the far-infrared. This issue has been previously investigated by a number of authors, and mitigation strategies based on AllWISE data release have been proposed to identify and remove affected sources \citep[e.g.][]{Koenig&Leisawitz2014,2016MNRAS.458.3479M,Marton2019,Silverberg2018ApJ...868...43S}. In particular, \citet[][see their eq. 1 to 4]{Koenig&Leisawitz2014} proposes that spurious detections can be separated by a series of cuts in terms of the signal-to-noise (\texttt{w?snr}) and profile fit reduced chi-squared (\texttt{w?chi2}) parameters in the AllWISE data release. 

Here, we adopted a strategy similar to  \citet{2016MNRAS.458.3479M,Marton2019}, where we built a Random Forest (RF) classifier to evaluate the reliability of WISE W3 and W4 detections based on AllWISE quality parameters and magnitudes. Nevertheless, we introduced two main modifications to \citet{Marton2019} approach. First, \citet{Marton2019} evaluated the reliability of W3 and W4 photometry simultaneously. Contrarily, we expect our sample to include sources that may only have one of these two bands above WISE's detection limits - thus we have trained one RF classifier for each photometric band. Second, \citet{Marton2019} classifier was trained with a sample of labelled AllWISE cutouts, where a researcher visually inspected the presence or lack of a source in image cutouts at a given band. Our initial tests with this approach revealed that such labels were highly subject to the researcher carrying out the labelling. To remove this subjectivity, we used instead Astropy's \texttt{DAOStarFinder} \citep{DAOPHOT,photoutils} to search the cutout images for local density maxima and label them in case a source detection was successful. We have built a training sample based on 5,000 AllWISE cutout images around known YSOs in Orion. These cutouts have been obtained for each of the W3 and W4 bands through the \citet{wise_cutouts} and covered a box of 10" around the positions of the sources. Our \texttt{DAOStarFinder} approach allowed us to label 2,356 and 1,970 sources as true detections in the W3 and W4 bands respectively. 

\begin{table}[bh]
\centering
\begin{tabular}{ l c c c }
    \hline\hline
     label  & precision & recall & f1-score \\
    \hline
    \multicolumn{4}{c}{W3} \\
    real   & 0.98 & 0.98 & 0.98 \\ 
    false   & 0.97 & 0.97 & 0.97 \\
    \hline

        \multicolumn{4}{c}{W4}\\
    real   & 0.90 & 0.91 & 0.90 \\
    false  & 0.90 & 0.89 & 0.90\\
    \hline
   \end{tabular}
   
\caption{Classification performance report of the Random Forest model for AllWISE's W3 (top) and W4 (bottom) bands.}\label{tab:class_w3w4}
\end{table}

\begin{figure}
\centering
\includegraphics[width=0.99\linewidth]{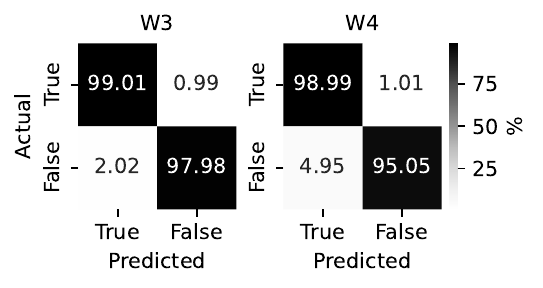}
\caption{Confusion matrices for the training for the RF classifier for AllWISE's W3 (left) and W4 (right). The top figure shows the confusion matrix for the W3 band, and the bottom figure shows the confusion matrix for the W4 band. }
\label{confusion_matrix}
\end{figure}

Our RF classifiers\footnote{We experimented with various machine learning classification methods, including also Logistic Regression, k-Neighbours, Naive Bayes, Decision Tree, Support Vector Machine, and Neural Networks, but found that the RF classifier yields the highest classification accuracy during the testing phase.} were trained on samples of 3,800 labelled sources (with 50-50\% labelled as true and false detections), using 21 features per band: AllWISE profile-fit magnitudes and their uncertainties (\texttt{w?mpro}, \texttt{w?sigmpro}),  signal-to-noise (\texttt{w?snr}) and profile fit reduced chi-squared (\texttt{w?chi2}), the magnitudes and uncertainties at 8 different apertures (\texttt{w?mag\_\#} and \texttt{w?sigm\_\#}, where \texttt{\#} refer to the 8 available apertures within AllWISE), and the ratios between magnitudes at consecutive apertures. Classification reports for both bands are given in Table \ref{tab:class_w3w4} and Confusion matrices are shown in Fig.~\ref{confusion_matrix}.

\begin{figure}
\includegraphics[width=0.7\linewidth]{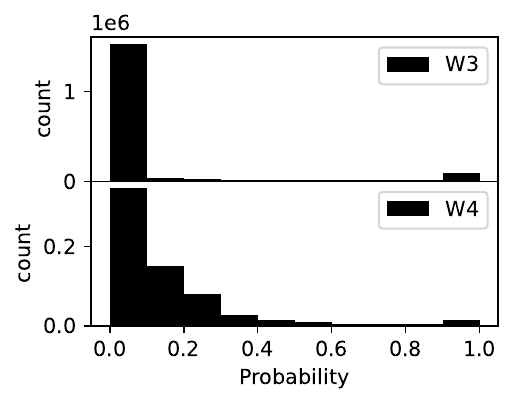}
\caption{Distribution of reliability probability of AllWISE W3 and W4 magnitudes for sources in the field of the OSFC.}
\label{prob_wise_hist}
\end{figure}

Next, we have employed the trained RF classifiers to estimate reliability probabilities for all available W3 and W4 AllWISE data in the OSFC field. This allowed us to flag 1,760,122 and 660,689 AllWISE observations in the W3 and W4 bands respectively. The distribution of reliability probabilities for this sample is shown in Fig.~\ref{prob_wise_hist}. For our purpose in this paper, we have removed all W3 and W4 data that had probability inferior to 60\%. Given the sources in our compilation with AllWISE counterparts, 17,453 (W3) and 6,581 (W4) sources had reliability probability derived, out of which 10,688 (W3) and 4,529 (W4) were under the threshold. Additionally, 404 (W3) and 230 (W4) sources had AllWISE counterparts in these bands but did not have all features used by the classifier available and we also eliminated these from the compilation. 

\section{$\alpha_{IR}$ reliability flags}\label{app:alpha_ir}

\begin{figure}
    \centering
       \includegraphics[width=0.95\linewidth]{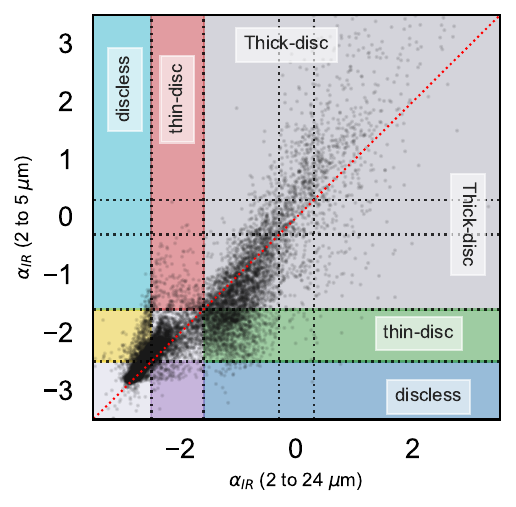}
\includegraphics[width=0.95\linewidth]{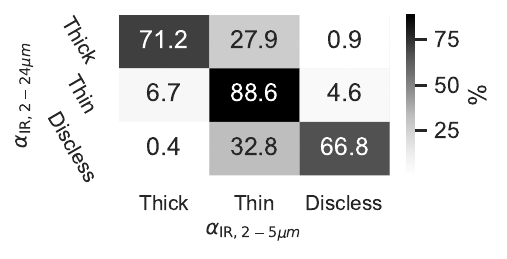}       
    \caption{Comparison of $\alpha_{IR}$-indices estimated with ($2-24\mu$m) and without ($2-5\mu$m) illustrating the effect of mid-infrared data availability into reported $\alpha_{IR}$-indices. Top: $\alpha_{IR,2-5\mu m}$ vs $\alpha_{IR,2-24\mu m}$. Bottom: Fraction of undersampled classification recovered as a function of disc class normalised by the number of sources in each class classified with the $\alpha_{IR,2-24\mu m}$ index.}
    \label{fig:alpha_ir_validation}
\end{figure}

 To quantify the effect of the reduced availability of mid/far infrared data on our $\alpha_{IR}$-indices (Sect.~\ref{sec:alphaindex}), we carried out an undersampling exercise. Recalling our requirement of two data points at least 2 $\mu m$
away in wavelength for $\alpha_{IR}$ derivation, we selected a testing sample of 16\,282 sources that followed this requirement for wavelengths $\lambda<5\mu$m, and had at least one additional data point in the range $5-24\mu$m. For this sample, we derived an additional $\alpha_{IR,2-5\mu m}$-index using only the data between $2-5\mu$m.  Fig.~\ref{fig:alpha_ir_validation} shows a comparison of this index to our main $\alpha_{IR}$. For the specific discussion in this Appendix, we limit our $\alpha_{IR}$ classes to three: discless, ($\alpha_{IR}\leq -2.5$)  thin disc ($-2.5<\alpha_{IR}< -1.6$) and thick disc ($\alpha_{IR}\leq -2.5$). We find that $\sim73.4\%$ of all sources had the same class with either $\alpha_{IR}$ definition (white and gray-shaded areas in Fig.~\ref{fig:alpha_ir_validation});  $\sim14.8\%$ sources were discless sources miss-detected as thin disc (yellow); $\sim8.6\%$ were thick discs miss-detected as thin disc (green);  $\sim1.6\%$ were thin discs miss-detected as thick disc (red); $\sim1.1\%$ of sources were thin discs miss-detected as discless (purple); $\sim0.3\%$ were thick discs miss-detected as discless (blue); and $\lesssim0.2\%$ sources were discless miss-detected as thick discs (cyan). With 33\% of our YSO candidates having no SED datapoints for $\lambda\geq5\mu$m, we expected such bias to affect $\sim9\%$ of our $\alpha_{IR}$-indices. We note that this estimation does not take into account domain-motivated choices of the wavelength range used for $\alpha_{IR}$, and we point the interested reader to \citet{2019AandA...622A.149G} for an in-depth discussion of the effect of $\alpha_{IR}$ definitions on disc classification. 
We also note that the numbers discussed so far refer to the total number of sources analysed. In the bottom plot of Fig.~\ref{fig:alpha_ir_validation}, we detailed miss-classification rates normalised by the number of sources in each class derived with $\alpha_{IR,2-24\mu m}$. Although only 71.2 and 66.9\% of the original thick discs and diskless sources are recovered as such, these classes have relatively large purity, with 93.7\% of thick disks and 97.8\% of discless sources correctly detected as such with the  $\alpha_{IR,2-5\mu m}$ index. In contrast, while 88.6\% of thin discs are recovered as such, this class has a much smaller purity of 74.8\%.

\section{Summary of studies included in the historical compilation}\label{app:historical_bib}

Tab.~\ref{tab:papers_with_data} presents a summary of peer-reviewed scientific publications with data collated into the NEMESIS catalogue of YSOs for the Orion Star Formation Complex.

\longtab[1]{

\begin{longtable}{p{0.37\linewidth}p{.34\linewidth}p{.025\linewidth}p{.025\linewidth}p{.025\linewidth}p{.025\linewidth}p{.025\linewidth}}
\caption{\label{tab:papers_with_data}List of peer-reviewed scientific publications with data collated as part of the historical data compilation behind the NEMESIS Catalogue of YSOs in the OSFC (Sect.~\ref{sec:historical-dtypes}). Data types are flagged as follows:\\ 
\texttt{Source list}: reference contributing the YSO source list with no further observable collected. \\
\texttt{SpT}: Spectral types collected in Sect.~\ref{sec:hrdiagram}.\\ 
\texttt{$T_\mathrm{eff/bol}$}: Effective or bolometric temperatures collected in Sect.~\ref{sec:hrdiagram}.\\ 
\texttt{SED}: Photometric data processed into SED in Sect.~\ref{sec:SED}. \\
 \texttt{disc}: Data types related to the infrared emission by circumstellar material collected in Sect.~\ref{sec:ir_emission}.\\
 \texttt{Li}: Observables related to the Lithium I line collected in Sect.~\ref{sec:lithium}.\\
 \texttt{logg}: logg measurements collected in Sect.~\ref{sec:gravity}. \\
 \texttt{EW$_\mathrm{gravity}$}: gravity-proxy emission lines collected in Sect.~\ref{sec:gravity}. \\
\texttt{$H\alpha$:} Observables related to the $H\alpha$ line collected in Sect.~\ref{sec:acc}.\\ 
\texttt{EW$_\mathrm{acc}$}: equivalent widths of accretion-related lines (except $H\alpha$) collected in Sect.~\ref{sec:acc}.\\  
\texttt{RV}: Radial velocity observations collected in Sect.~\ref{sec:kinematic}.\\
\texttt{rotation}: observables related to stellar rotation collected in Sect.~\ref{sec:spin}.\\
\texttt{X-ray}: observables related to the X-ray emission of YSOs collected in Sect.~\ref{sec:xray}.\\
\texttt{bin}: Multiplicity information collected in Sect.~\ref{sec:multiplicity};\\ 
 How the tabulated data was retrieved (Sect.~\ref{sec:histDataRetrival}):\\
 \texttt{CDS}: through the Strasbourg astronomical Data Center;\\
 \texttt{pub}: from the publisher's webpage;\\
 {\texttt{man}}: extracted manually directly from the manuscript;\\
 {\texttt{TbC}} extracted using TableConvert:\\
 {\texttt{Tba}}: extracted using Tabula.}\\
 \hline\hline
    Reference & {Data type}  & \texttt{CDS} & \texttt{pub} & {\texttt{man}} & {\texttt{TbC}} & {\texttt{Tba}} \\
\hline
\endfirsthead
\caption{continued.}\\
\hline\hline
    Reference & {Data type}  & \texttt{CDS} & \texttt{pub} & {\texttt{man}} & {\texttt{TbC}} & {\texttt{Tba}} \\
\hline
\endhead
\hline
\endfoot
    \citet{1945ApJ...102...74S}   &  bin  &  &  & x &  & \\
    \citet{1976ApJ...205..462P}   &  bin  &  &  & x &  & \\
    \citet{1976Natur.262..116H}   &  bin  &  &  & x &  & \\
    \citet{1988AJ.....96.1956M}   &  bin  &  &  & x &  & \\
    \citet{1989AandA...222..117B} &  bin  &  &  & x &  & \\
    \citet{1991AJ....101.2184M}    &  bin  &  &  & x &  & \\
    \citet{1991ApJ...367..155A}    &  bin  &  &  & x &  & \\
    \citet{1993PASP..105..721P}    &  bin  &  &  & x &  & \\
    \citet{1994ApJ...437..361G}    &  X-ray  & x &  &  &  & \\
    \citet{1995AandAT....8..249K}  &  SpT, EW$_\mathrm{acc}$, H$\alpha$  &  &  & x &  & x\\
    \citet{1996ApJ...469..884G}    &  bin  &  &  & x &  & \\
    \citet{1997AJ....113.1733H}    &  $T_\mathrm{eff/bol}$, SED  & x &  &  &  & \\
    \citet{1998AandA...337..183B}  &  bin  &  &  & x &  & \\
    \citet{1998AJ....116.1816H}    &  EW$_\mathrm{acc}$ disc, SED,  & x &  &  &  & \\
    \citet{1999AJ....117.2941S}    &  Li, RV, H$\alpha$, SED  & x &  &  &  & \\
    \citet{1999IBVS.4809....1L}    &  bin  &  &  & x &  & \\
    \citet{2000AandA...353..186A}  &  X-ray, Li, rotation, RV, SpT, H$\alpha$, bin  &  &  & x & x & x\\
    \citet{2000AandA...361L..49C}  &  bin  &  &  & x &  & \\
    \citet{2000AJ....119..261H}   &  rotation, EW$_\mathrm{acc}$  &  & x & x &  & \\
    \citet{2000AJ....119.3026R}   &  SpT, SED  & x &  &  &  & \\
    \citet{2001AandA...375..130C} &  bin  &  &  & x &  & \\
    \citet{2001AJ....121.1676R}  &  rotation, SpT, SED  & x &  &  &  & \\
    \citet{2001AJ....121.2124D}  &  Li, RV, H$\alpha$, bin, SED  & x &  &  &  & \\
    \citet{2001AJ....121.3160C}  &  rotation, SED  & x &  &  &  & \\
    \citet{2001AJ....122.3258R}  &  rotation, bin  &  & x &  &  & \\
    \citet{2001ApJ...553..299P}  &  bin  &  &  & x &  & \\
    \citet{2001ApJ...556..830B}  &  SpT, SED  &  & x &  &  & \\
    \citet{2001KFNT...17..409K}  &  SpT, bin  & x &  &  &  & \\
    \citet{2002AandA...396..513H} &  rotation  & x &  &  &  & \\
    \citet{2002AJ....124.1089M} &  bin  &  &  & x &  & \\
    \citet{2002ApJ...574..258F} &  X-ray  & x &  &  &  & \\
    \citet{2003AandA...402..963M} &  Li, SpT, H$\alpha$, SED  & x &  & x & x & \\
    \citet{2003ApJ...582..382F} &  X-ray  & x &  &  &  & \\
    \citet{2003ApJ...598..375S} &  X-ray  & x &  &  &  & \\
    \citet{2004AandA...416..677A} &  Li, RV, $T_\mathrm{eff/bol}$, SpT, H$\alpha$, bin  &  & x &  & x & \\
    \citet{2004AandA...419..249S} &  rotation $T_\mathrm{eff/bol}$, SpT, logg, EW$_\mathrm{acc}$, H$\alpha$, SED  & x & x & x & x & \\
    \citet{2004AJ....128..787R} &  X-ray  & x &  &  &  & \\
    \citet{2004AJ....128.2316S} & SED  & x &  &  &  & \\
    \citet{2004ApJ...610.1064B} &  SpT, H$\alpha$, SED  & x &  &  &  & \\
    \citet{2004ApJ...611..940S} &  X-ray  &  & x &  & x & \\
    \citet{2004ApJ...613..374A} &  disc, SED  &  &  & x & x & \\
    \citet{2004ApJS..151..357S} &  Li, bin  &  &  & x &  & \\
    \citet{2004MNRAS.353..697B} &  bin  &  &  & x &  & \\
    \citet{2005AandA...429.1007S} &  rotation, SED  & x & x &  & x & \\
    \citet{2005AJ....129..363S} &  Li, rotation, RV, SpT, H$\alpha$, bin  & x &  &  &  & \\
    \citet{2005AJ....129..907B} &  Li, $T_\mathrm{eff/bol}$, SpT, H$\alpha$  & x &  &  &  & \\
    \citet{2005AJ....129.1534R} &  bin, SED  & x &  &  &  & \\
    \citet{2005AJ....130.1763S} &  disc, SED  & x &  &  &  & \\
    \citet{2005ApJS..160..319G} &  X-ray  & x &  &  &  & \\
    \citet{2005ApJS..160..353G} & Source list  &  &  &  &  & \\
    \citet{2005MNRAS.356...89K} &  Li, RV, EW$_\mathrm{gravity}$, bin  & x &  & x &  & \\
    \citet{2005MNRAS.356.1583B} &  RV, EW$_\mathrm{gravity}$, SED  & x &  &  &  & \\
    \citet{2006AandA...446..501F} &  X-ray, SpT  & x &  &  &  & \\
    \citet{2006AandA...458..461K} &  bin  & x &  &  &  & \\
    \citet{2006AJ....132.1763H} &  bin  &  &  & x &  & \\
    \citet{2006Natur.440..311S} &  bin  &  &  & x &  & \\
    \citet{2007AandA...466..917C} &  RV, SpT, SED & x &  &  &  & \\
    \citet{2007AandA...468L...5B} &  bin  &  &  & x &  & \\
    \citet{2007AJ....134.2272R} &  SpT, bin  & x &  &  &  & \\
    \citet{2007ApJ...657..884L} &  SpT, H$\alpha$  & x &  & x &  & \\
    \citet{2007ApJ...661.1119B} &  Li, RV, H$\alpha$  & x &  &  &  & \\
    \citet{2007ApJ...662.1067H} &  disc, SED  & x &  &  &  & \\
    \citet{2007ApJ...664..481B} &  disc, SED  & x &  &  &  & \\
    \citet{2007ApJ...671.1784H} &  disc, SED  & x & x &  &  & \\
    \citet{2007MNRAS.380..541I} &  bin  &  &  & x &  & \\
    \citet{2008AandA...478..667C} & SED & x &  &  &  & \\
    \citet{2008AandA...481..747S} &  bin  &  &  & x &  & \\
    \citet{2008AandA...485..931C} &  SpT, bin, SED & x &  &  &  & \\
    \citet{2008AandA...488..167S} &  Li, rotation, RV, SpT, EW$_\mathrm{acc}$, H$\alpha$, bin  & x &  &  &  & \\
    \citet{2008AandA...491..961L} &  X-ray  &  & x &  & x & \\
    \citet{2008AandA...492..277B} &  $T_\mathrm{eff/bol}$  & x &  &  &  & \\
    \citet{2008ApJ...674..329C} &  bin  &  &  & x &  & \\
    \citet{2008ApJ...676.1109F} &  RV, H$\alpha$  & x &  &  &  & \\
    \citet{2008ApJ...688..362L} & SED  & x &  &  &  & \\
    \citet{2008MNRAS.385.2210M} &  RV, EW$_\mathrm{gravity}$, bin, SED,  & x &  &  &  & \\
    \citet{2009AandA...502..883R} &  rotation, SED,  & x &  &  &  & \\
    \citet{2009AandA...504..461F} &  Li, SpT, EW$_\mathrm{acc}$, H$\alpha$, disc, SED,  & x & x &  &  & \\
    \citet{2009AandA...508.1301B} &  rotation, RV, SpT, H$\alpha$,  &  & x & x & x & \\
    \citet{2009AandA...508.1313F} &  rotation, $T_\mathrm{eff/bol}$, SpT, SED,  & x &  &  &  & \\
    \citet{2009AJ....137.3487M} &  bin,  &  &  & x &  & \\
    \citet{2009ApJ...697..493B} &  bin,  &  &  & x &  & \\
    \citet{2009ApJ...697..713M} &  bin,  &  &  & x &  & \\
    \citet{2009ApJ...697.1103T} &  RV, bin, SED,  & x &  &  &  & \\
    \citet{2009ApJ...706..896M} &  disc, SED,  & x &  &  &  & \\
    \citet{2009ApJ...707..705H} &  disc, SED,  & x &  &  &  & \\
    \citet{2009ApJS..183..261D} &  SpT, H$\alpha$, SED,  & x &  &  &  & \\
    \citet{2009MNRAS.400..354A} &  bin,  &  &  & x &  & \\
    \citet{2009MNRAS.400..603P} &  rotation, H$\alpha$, SED,  & x &  &  &  & \\
    \citet{2010AJ....140.1214C} &  SpT, EW$_\mathrm{acc}$, veiling, bin, disc, SED,  & x &  &  &  & \\
    \citet{2010Ap.....53..367G} &  bin,  &  &  & x &  & \\
    \citet{2010ApJ...722.1092D} &  $T_\mathrm{eff/bol}$, SpT,  & x &  &  &  & \\
    \citet{2010ApJS..191..389C} &  rotation, SED,  & x &  &  &  & \\
    \citet{2010MNRAS.401.2739L} &  bin,  &  &  & x &  & \\
    \citet{2011AandA...525A..47R} &  $T_\mathrm{eff/bol}$, H$\alpha$, disc, SED,  & x &  & x &  & \\
    \citet{2011AandA...526A..21B} &  X-ray, disc, SED,  & x &  &  &  & \\
    \citet{2011AandA...530A.150F} &  X-ray, SpT, disc,  & x &  &  &  & \\
    \citet{2011AandA...536A..63B} &  Li, $T_\mathrm{eff/bol}$, SpT, EW$_\mathrm{gravity}$,  & x &  &  &  & \\
    \citet{2011AJ....141..127I} &  X-ray, SpT, H$\alpha$,  &  & x & x &  & \\
    \citet{2011AJ....142...60V} &  bin,  & x &  &  &  & \\
    \citet{2011ApJ...733...50M} &  rotation,  &  & x &  &  & \\
    \citet{2011ApJ...743...64B} &  disc, SED,  & x &  &  &  & \\
    \citet{2012AandA...540A..46D} &  bin,  &  &  & x &  & \\
    \citet{2012AandA...546A..59C} &  Li, SpT, EW$_\mathrm{acc}$, H$\alpha$, disc,  &  &  & x & x & \\
    \citet{2012AandA...547A..80B} &  rotation, SpT, EW$_\mathrm{acc}$, H$\alpha$, disc,  & x &  & x &  & \\
    \citet{2012AandA...548A..56R} &  SpT, EW$_\mathrm{acc}$, H$\alpha$,  & x &  &  &  & \\
    \citet{2012AJ....144...31K} &  disc, SED,  & x &  &  &  & \\
    \citet{2012AJ....144..192M} &  disc, SED,  & x &  &  &  & \\
    \citet{2012ApJ...745...58G} &  bin,  &  &  & x &  & \\
    \citet{2012ApJ...748...14D} &  $T_\mathrm{eff/bol}$,  & x &  &  &  & \\
    \citet{2012ApJ...752...59H} &  Li, SpT, EW$_\mathrm{acc}$, H$\alpha$, SED,  & x &  &  &  & \\
    \citet{2012ApJ...753..149M} &  rotation, SpT, bin,  & x &  &  &  & \\
    \citet{2012ApJ...754...30P} & SED,  & x &  &  &  & \\
    \citet{2012ApJ...755..154M} &  H$\alpha$, disc,  & x &  &  &  & \\
    \citet{2013AandA...557A..63C} &  bin,  &  &  & x &  & \\
    \citet{2013AJ....146...85H} &  SpT, bin,  & x &  &  &  & \\
    \citet{2013ApJ...764..114H} &  Li, RV, SpT, disc, SED,  & x &  &  &  & \\
    \citet{2013ApJ...767...36S} &  $T_\mathrm{eff/bol}$, bin, disc, SED,  & x &  &  &  & \\
    \citet{2013ApJ...768...67W} &  bin,  &  &  & x &  & \\
    \citet{2013ApJ...768...99P} &  X-ray,  & x &  &  &  & \\
    \citet{2013ApJ...769..149K} &  X-ray, $T_\mathrm{eff/bol}$, SpT, bin, disc,  & x &  &  &  & \\
    \citet{2013ApJS..207....5F} &  Li, SpT, EW$_\mathrm{acc}$, H$\alpha$, disc, SED,  & x &  &  &  & \\
    \citet{2013ApJS..208...28S} &  H$\alpha$, SED,  & x &  &  &  & \\
    \citet{2013ApJS..209...29K} & SED,  & x &  &  &  & \\
    \citet{2013ApJS..209...31P} & Source list,  &  &  &  &  & \\
    \citet{2013ApJS..209...32B} &  X-ray, disc,  & x &  &  &  & \\
    \citet{2013MNRAS.429..775L} &  X-ray,  & x &  &  &  & \\
    \citet{2013MNRAS.434..966S} &  Li, RV, veiling, H$\alpha$,  & x &  &  &  & \\
    \citet{2014AandA...564A..29B} & SED,  & x &  &  &  & \\
    \citet{2014AandA...570A..30P} &  H$\alpha$, bin, SED,  & x &  &  &  & \\
    \citet{2014AandA...570A.118F} &  bin,  &  &  & x &  & \\
    \citet{2014AandA...572A..89S} &  $T_\mathrm{eff/bol}$, SED,  & x &  &  &  & \\
    \citet{2014ApJ...782....8I} &  $T_\mathrm{eff/bol}$, SpT,  & x &  &  &  & \\
    \citet{2014ApJ...787..107K} &  SpT, bin,  & x &  &  &  & \\
    \citet{2014ApJ...794...36H} &  Li, RV, SpT, EW$_\mathrm{gravity}$, H$\alpha$, bin, disc, SED,  & x &  & x & x & \\
    \citet{2014ApJ...794..146T} &  RV, $T_\mathrm{eff/bol}$, SpT, EW$_\mathrm{gravity}$, EW$_\mathrm{acc}$, H$\alpha$,  & x &  &  &  & \\
    \citet{2014MNRAS.444.1793D} &  SpT, H$\alpha$,  & x &  &  &  & \\
    \citet{2015AandA...581A.140S} &  disc, SED,  & x &  &  &  & \\
    \citet{2015AJ....150..100K} &  SpT, EW$_\mathrm{acc}$, H$\alpha$, disc, SED,  & x &  &  &  & \\
    \citet{2015AJ....150..132R} &  rotation, bin, SED,  & x &  &  &  & \\
    \citet{2015MNRAS.450.3490D} &  $T_\mathrm{eff/bol}$, SpT, H$\alpha$, disc, SED,  & x &  &  &  & \\
    \citet{2016AandA...587A.153M} & SED,  & x &  &  &  & \\
    \citet{2016AandA...593A...7H} &  SpT, disc,  & x &  &  &  & \\
    \citet{2016AJ....151....5M} &  disc, SED,  &  & x &  &  & \\
    \citet{2016AJ....152..198K} &  rotation,  & x &  &  &  & \\
    \citet{2016ApJ...818...59D} &  rotation, $T_\mathrm{eff/bol}$, logg, SED,  & x &  &  &  & \\
    \citet{2016ApJ...821....8K} &  Li, RV, $T_\mathrm{eff/bol}$, bin,  & x &  &  &  & \\
    \citet{2016ApJ...821...52K} &  bin, disc,  & x &  &  &  & \\
    \citet{2016ApJ...825...91L} &  bin, disc,  & x &  &  &  & \\
    \citet{2016ApJS..224....5F} &  $T_\mathrm{eff/bol}$, disc, SED,  & x &  &  &  & \\
    \citet{2016ApJS..226....8K} &  Li, $T_\mathrm{eff/bol}$, SpT, H$\alpha$, bin, disc,  & x &  &  &  & \\
    \citet{2016MNRAS.457.3372M} &  bin,  &  &  & x &  & \\
    \citet{2017AandA...608L...2P} &  X-ray,  &  & x &  & x & \\
    \citet{2017AJ....153..188F} &  Li, SpT, H$\alpha$, disc,  & x &  &  &  & \\
    \citet{2017AJ....154...14S} &  Li, $T_\mathrm{eff/bol}$, SpT, H$\alpha$, disc,  & x &  &  &  & \\
    \citet{2017AJ....154...29K} &  Li, SpT, H$\alpha$,  & x &  &  &  & \\
    \citet{2017ApJ...834..142K} &  bin,  & x &  &  &  & \\
    \citet{2017ApJ...841...95S} &  bin,  &  &  & x &  & \\
    \citet{2017ApJ...844..138K} &  Li, RV, $T_\mathrm{eff/bol}$, H$\alpha$, disc,  & x &  &  &  & \\
    \citet{2017ApJ...851...14J} &  bin,  &  & x &  &  & \\
    \citet{2017ApJS..229...28G} &  X-ray, SED,  & x &  &  &  & \\
    \citet{2017MNRAS.468..931M} & SED,  & x &  &  &  & \\
    \citet{2017PASP..129h4201F} &  bin,  & x &  &  &  & \\
    \citet{2018AandA...620A.116G} &  bin,  &  &  & x &  & \\
    \citet{2018AandA...620A.128V} &  $T_\mathrm{eff/bol}$, H$\alpha$, bin, disc, SED,  & x &  &  &  & \\
    \citet{2018AandA...620A.172Z} &  RV,  & x &  &  &  & \\
    \citet{2018AJ....156...84K} &  RV, $T_\mathrm{eff/bol}$, logg,  & x &  &  &  & \\
    \citet{2018AN....339...60C} &  bin,  &  &  & x &  & \\
    \citet{2018ApJ...863...13G} &  SpT, disc, SED,  & x &  &  &  & \\
    \citet{2018ApJ...869...72Y} &  $T_\mathrm{eff/bol}$, logg, EW$_\mathrm{acc}$, disc, SED,  & x &  &  &  & \\
    \citet{2018ApJS..236...27C} &  disc, SED,  & x &  &  &  & \\
    \citet{2018MNRAS.474.5406D} &  bin,  &  &  & x &  & \\
    \citet{2018MNRAS.477..298G} &  bin,  & x &  &  &  & \\
    \citet{2018MNRAS.477..298G} &  disc,  &  &  &  &  & \\
    \citet{2018MNRAS.477.3145J} &  rotation,  &  & x &  &  & \\
    \citet{2018MNRAS.478.1825D} &  bin,  &  &  & x &  & \\
    \citet{2019AandA...622A.149G} &  disc, SED,  & x &  &  &  & \\
    \citet{2019AandA...627A..57J} &  bin,  & x &  &  &  & \\
    \citet{2019AandA...629A.114C} &  Li, SpT, EW$_\mathrm{acc}$, H$\alpha$,  & x &  & x & x & \\
    \citet{2019AJ....157...85B} &  Li, $T_\mathrm{eff/bol}$, SpT, EW$_\mathrm{gravity}$, H$\alpha$, SED,  & x &  &  &  & \\
    \citet{2019AJ....157..109K} &  bin, SED,  & x &  &  &  & \\
    \citet{2019AJ....157..196K} &  RV, $T_\mathrm{eff/bol}$, logg, veiling, bin, disc,  & x &  &  &  & \\
    \citet{2019ApJ...871...72M} &  bin,  &  &  & x &  & \\
    \citet{2019ApJ...884....6M} &  RV,  & x &  &  &  & \\
    \citet{2019ApJ...886....6T} &  bin,  &  &  & x &  & \\
    \citet{2019ApJ...886...95D} &  bin,  &  &  & x &  & \\
    \citet{2019MNRAS.486.1718S} & SED,  & x &  &  &  & \\
    \citet{2020AJ....160...86B} &  bin,  &  &  & x &  & \\
    \citet{2020AJ....160..268T} &  bin,  & x &  &  &  & \\
    \citet{2020AJ....160..279K} & Source list,  &  &  &  &  & \\
    \citet{2020ApJ...893...56M} &  SpT, H$\alpha$,  & x &  &  &  & \\
    \citet{2020ApJ...896...79R} &  $T_\mathrm{eff/bol}$,  & x &  &  &  & \\
    \citet{2020ApJ...896...81S} &  bin,  & x &  &  &  & \\
    \citet{2020ApJ...905..119F} & SED,  & x &  &  &  & \\
    \citet{2020MNRAS.496.4701J} &  RV, $T_\mathrm{eff/bol}$, logg, SED,  & x &  &  &  & \\
    \citet{2020MNRAS.497..632L} &  bin,  &  &  & x &  & \\
    \citet{2021AandA...656A.138F} & Source list,  &  &  &  &  & \\
    \citet{2021AJ....162...90P} &  rotation, RV, $T_\mathrm{eff/bol}$, SpT, bin, disc,  &  & x &  & x & \\
    \citet{2021ApJ...908...49F} &  SpT, veiling, bin,  &  & x &  &  & \\
    \citet{2021ApJ...911..153H} &  disc,  & x &  &  &  & \\
    \citet{2021ApJ...921..110L} &  EW$_\mathrm{acc}$,  & x & x & x &  & \\
    \citet{2021ApJ...923..177S} &  rotation, $T_\mathrm{eff/bol}$, bin, SED,  & x &  &  &  & \\
    \citet{2021MNRAS.501.1243A} &  bin,  &  &  & x &  & \\

    \citet{2021MNRAS.506.4232K} &  Li, rotation, RV, $T_\mathrm{eff/bol}$, logg,  &  &  &  &  & \\
    \citet{2021RNAAS...5...36P} &  bin,  &  &  & x &  & \\
    \citet{2022AandA...659A..85F} &  RV, $T_\mathrm{eff/bol}$,  & x &  &  &  & \\
    \citet{2022AandA...666A..55F} &  $T_\mathrm{eff/bol}$,  & x &  &  &  & \\
    \citet{2022AJ....163...74T} &  $T_\mathrm{eff/bol}$, SpT, EW$_\mathrm{acc}$, H$\alpha$, bin,  & x &  &  &  & \\
    \citet{2022AJ....164..137K} &  rotation,  & x &  &  &  & \\
    \citet{2022AJ....164..201P} &  rotation,  & x &  &  &  & \\
    \citet{2022ApJ...924...84C} &  $T_\mathrm{eff/bol}$, disc,  &  & x &  &  & \\
    \citet{2022ApJ...925..112D} &  bin,  &  &  & x &  & \\
    \citet{2022ApJ...926..141T} &  rotation, RV, $T_\mathrm{eff/bol}$, veiling, bin,  & x &  &  &  & \\
    \citet{2022ApJ...941..161D} &  bin,  &  & x &  &  & \\
    \citet{2023AJ....165..205H} &  Li, SpT, H$\alpha$,  & x &  &  &  & \\
    \citet{2023ApJ...944...49F} &  $T_\mathrm{eff/bol}$, disc,  &  & x &  &  & \\
    \citet{2023ApJ...951..139D} &  $T_\mathrm{eff/bol}$, SpT, SED,  &  & x &  &  & \\
    \citet{2023MNRAS.523..169S} &  rotation, $T_\mathrm{eff/bol}$, disc,  &  & x &  &  & \\
\hline
    \end{longtable}
}

\section{Database Summary}\label{app:DBsummary}

The database curated as part of the NEMESIS Catalogue of YSOs for the OSFC can be accessed via SQL (at \url{https://www.astro.unige.ch/nemesis/}) and through CDS. The 19 thematic tables with data compiled in the present study are organised in 12 directories. A full description of directories, sub-tables, column names, units and data availability is also available as part of the online material (\texttt{nemesis\_osfc\_description.csv}). This appendix further summarises key data products, the number of individual measurements collected for each observable, and the number of sources with at least one measurement. We stress that the data is organised following a relational database paradigm.  Each source is associated with a unique \texttt{NEMESIS\_ID} key. This unique source identifier should be used to group multiple measurements (when available) for a given source and match different thematic tables.

For each quantity described in Tabs.~\ref{app:tab:main} to \ref{app:tab:xray}, when available, the catalogue includes the following accompanying fields: 

\begin{itemize}
    \item \texttt{\#}: reported quantity, for example \texttt{Teff} for $T_\mathrm{eff}$ or \texttt{EW\_Ha} for equivalent width of the H$\alpha$ line;
    \item \texttt{\#\_ref}: NASA/ADS \texttt{bibcode} for the original paper linked to the reported quantity \texttt{\#}; 
    \item \texttt{e\_\#} uncertainty;     \item \texttt{e\_up\_\#}/\texttt{e\_low\_\#} upper and lower uncertainty values;
    \item \texttt{f\_\#} flag indicating that a limit value is reported;
    \item \texttt{\#\_comment}: String reporting author's comments from the original paper. 
\end{itemize}

\begin{description}
    \item Tab.~\ref{app:tab:main}: \texttt{main/}
    \begin{itemize}
        \item \texttt{id.csv}: Complete list of sources in the NEMESIS YSO Catalogue for the OSFC, along with their reference coordinates.
        \item  \texttt{contamination.csv}: Contamination flags (Sects.~\ref{sec:duplicate} and \ref{sec:contamination}).
        \item \texttt{simbad.csv}: List of possible counterparts in the Simbad database (Sect.~\ref{sec:comparison-simbad}).
    \end{itemize}
    \item Tab.~\ref{tab:hr_diagram}: \texttt{hrd/} Quantities related to the HRD and BLT-diagrams (Sect.~\ref{sec:hrdiagram});
    \begin{itemize}
        \item \texttt{temperature.csv}: temperatures (Sect.~\ref{sec:Teff} and \ref{sec:Tbol}.
        \item \texttt{spt.csv} Spectral types (Sect.~\ref{sec:SpT}).
    \end{itemize}
    \item  Tab.~\ref{tab:SED}: \texttt{sed/} SED database (Sect.~\ref{sec:SED})
    \begin{itemize}
        \item \texttt{sed\_summary.csv} Sumarry of data availability for SEDs
        \item \texttt{nemesis/} directory with SEDs compiled in the present study (Sect. \ref{sec:SED}).
        \item \texttt{vizier/} directory with SEDs compiled with the VizieR Photometry Viewer tool (Sect.~\ref{sec:VizierSED}).
    \end{itemize}
    
    \item Tab.~\ref{tab:app:disc} \texttt{ir/} : Quantities related to IR YSO classification schemes discussed Sect.~\ref{sec:ir_emission};
        \begin{itemize}
        \item \texttt{standard\_classification\_nemesis.csv} $\alpha_{IR}$ YSO classification derived in this study.
        \item \texttt{disk\_literature.csv}: IR-based YSO classification collected from the literature.
    \end{itemize}
    \item Tab.~\ref{tab:Lithium} \texttt{lithium/li.csv}: Lithium observables (Sect.~\ref{sec:lithium}); 
    \item Tab.~\ref{app:tab:gravity} \texttt{gravity/} : gravity observables (Sect.~\ref{sec:gravity});
        \begin{itemize}
        \item \texttt{ew\_gravity.csv}: gravity sensitive emission lines;
        \item \texttt{logg.csv}: $\log{g}$ measurements.
    \end{itemize}
    \item Tab.~\ref{tab:acc}: \texttt{accretion/} : Quantities related to accretion or material inflow/outflow (Sect.~\ref{sec:acc});
        \begin{itemize}
        \item \texttt{ha.csv}: Measurements related to the H$\alpha$ line (Sect.~\ref{sec:hydrogen_lines});
        \item \texttt{ew\_accretion\_other.csv}: measurement of diverse emission lines related to accretion (Sect.~\ref{sec:otherEW});
        \item \texttt{veiling.csv}: veiling measurements (Sect.~\ref{sec:veiling}).
    \end{itemize}
    \item Tab.~\ref{tab:RV} \texttt{kinematics/rv.csv} Radial velocity measurements (Sect.~\ref{sec:kinematic});
    \item Tab.~\ref{tab:var}  \texttt{variability/amplitude.csv} : variability amplitudes (Sect.~\ref{sec:var}).    
    \item Tab.~\ref{tab:Rot_data_types} \texttt{rotation/rotation.csv}:  quantities related to rotation of YSOs (Sect.~\ref{sec:spin});
    \item Tab.~\ref{app:tab:xray}: quantities related to the X-ray emission of YSOs (Sect.~\ref{sec:xray});
    \item Tab.~\ref{app:bin} \texttt{multiplicity/multiplicity\_label.csv}: Multiplicity labels (Sect.~\ref{sec:multiplicity}).    
\end{description}

\begin{table*}[]
    \caption{Summary of data types in the main table.
    \tablefoottext{a}{RA, Dec reported are in J2000 and refer to the first study adding the source to this compilation.}
       \tablefoottext{b}{When multiple possible counterparts exist within the Simbad database, this field inform how many possibilities there are.}
           \tablefoottext{c}{Labelled if included in the external catalogues/samples discussed in Sect.~\ref{app:RFlabels} following the values: \texttt{1}. GDR3 Galaxy high-purity sample; \texttt{2}. GDR3 QSO high-purity sample; \texttt{3}. MillionQua catalogue; \texttt{4}. MANGROVE catalogue; \texttt{5}. HyperLEDA database; \texttt{6}. GalaxyZoo catalogue; \texttt{-1}. further stellar sample; \texttt{-2}. giant sample; \texttt{-3}. at Orion (distance) sample; \texttt{-4}. YSO sample with spectroscopic support}
           \label{app:tab:main}}
 
    \centering
    \begin{tabular}{llcrr}
     \hline\hline
        key & description & units & count & \# sources\\
   \hline
    \multicolumn{5}{c}{\underline{Identification} (\texttt{main/id.csv})}\\
\texttt{RA},\texttt{DE}\tablefootmark{a}  & Source's right Ascension and Declination &  degree & 27\,879 & 27\,879 \\
   \texttt{Internal\_ID}  & NEMESIS Catalogue Internal ID & - & 27\,879 & 27\,879 \\
   
    \multicolumn{5}{c}{{\underline{Simbad counterparts} (\texttt{main/simbad.csv})}}\\
    
\texttt{main\_id} & Main identifier in Simbad &   - & 33\,846 & 25\,902 \\
\texttt{otype}  & Main Object Type in Simbad & - & 33\,690& 25\,901  \\
\texttt{otypes} & Full list of Object types in Simbad & - & 33\,615 & 25\,902 \\
  \texttt{ra\_simbad}, \texttt{dec\_simbad} & RA, Dec coordinates in Simbad& degree & 33\,846 & 25\,902  \\
    \texttt{multiple\_counterparts}\tablefootmark{b}  & number of possible counterparts & - & 12\,835 & 4\,891 \\
 \texttt{counterpart\_groupID} & identifier for grouping multiple possible counterparts &  - & 33\,846 & 25\,902  \\    
      \texttt{sep2simbad} & Separation to Simbad counterpart & arcsec & 12\,835 & 4\,891 \\
     \multicolumn{5}{c}{{\underline{Contamination} (\texttt{main/contamination.csv}))}}\\
   \texttt{massive\_star} & Flag indicative of likely massive star (Sect.~\ref{sec:massivestars}) & - & 80 & 80 \\
   \texttt{ms\_contaminant} & Likely MS contaminant(Sect.~\ref{sec:ms_contamination})& - & 2\,845 & 2\,845 \\

      \texttt{extraglactic\_prob} &  Probability of source being extragalactic (App.~\ref{app:galaxy})& - & 15\,855 & 15\,855 \\
      \texttt{extraglactic\_pred} &  Extragalactic predicted label (App.~\ref{app:galaxy}) & - & 15\,855 & 15\,855       \\
      \texttt{extraglactic\_label} & Labelled extragalactic contaminant  (Sect.~\ref{sec:galaxy} and App.~\ref{sec:GalLit}) & - & 125 & 125 \\
      \texttt{giant\_prob} & Probability of source being a Giant (App.~\ref{app:galaxy})& - & 20\,546 & 20\,546 \\
      \texttt{giant\_prob} & Probability of source being a Giant (App.~\ref{app:galaxy})& - & 20\,546 & 20\,546 \\
      \texttt{giant\_label}  & Labelled Giant contaminant (App.~\ref{sec:GalLit} and Sect.~\ref{sec:giants})& - & 490 & 490 \\      
      \texttt{strat\_label} & Labelled source in App.~\ref{app:galaxy}\tablefootmark{c} & - &  9\,529 &9\,529 \\
   
      \hline
\end{tabular}
\end{table*}

\begin{table*}[]
    \caption{Summary of data types related to HR-diagrams.
    \tablefoottext{a}{\texttt{SpT-PHOT} combined Spectral Typing and photometric data; \texttt{SED} based on SED model-fit; \texttt{SPEC} spectra-library fit to spectra; \texttt{PHOT} fit of photometric data to models using MCMC or Bayasean methods.}
    \tablefoottext{b}{Spectral types may be reported as a range with two MK SpTs separated by a symbol \texttt{-}. Unresolved multiple systems may have each components' spectral types reported separated by a symbol \texttt{+}.}
    \tablefoottext{c}{Extra information on the spectral typing reported by the original study. For example, luminosity classes are reported if available and emission spectra will include a flag \texttt{e}.}
    \label{tab:hr_diagram} }
    \centering
    \begin{tabular}{llcrr}
    \hline\hline
        key & description & units & count & \# sources\\
    \hline
    \multicolumn{5}{c}{\underline{Temperature} (\texttt{hrd/temperature.csv})}\\
  \texttt{Teff} & Effective temperature  & K & 73\,808 & 17\,589 \\
 \texttt{Teff\_type} &  Effective Temperature derivation method\tablefootmark{a}  & - & 74\,508 & 17\,901\\
 \texttt{Tbol} & Bolomnetric temperature & K & 700 & 381 \\
     \multicolumn{5}{c}{\underline{Spectral Type} (\texttt{hrd/spt.csv})}\\
  \texttt{SpT} & MK Spectral Types\tablefootmark{b}  & -  & 22\,504 & 11\,420 \\
  \texttt{SpT\_ext} & Extra spectral typing information reported\tablefootmark{c} & - & 3\,337 & 2\,756 \\
    \texttt{phot\_SpT} & \texttt{True} if Spectral typing was based on narrow-band photometry. & - & 263 & 263 \\\hline
\end{tabular}
\end{table*}

\begin{table*}[]
    \caption{Summary of the SED database (Sect.~\ref{sec:SED}). 
    \tablefoottext{a}{One \texttt{.csv} per source.}
    \tablefoottext{b}{\texttt{\#\#\#\#\#} refers to the catalogue \texttt{Internal\_ID} (see Tab.~\ref{app:tab:main}). \texttt{??} refers to: \texttt{nemesis} if SED was compiled as part of the historical data compilation (Sect.~\ref{sec:data-historical}) with complementation from large photometric surveys (Sect.~\ref{sec:largePhotSurveys}); \texttt{vizier} if SED was built with the Vizier-SED tool (Sect.~\ref{sec:VizierSED}. }
    \label{tab:SED}}
    \centering
    \begin{tabular}{rlc}
    \hline\hline
   field & description & units\\
   \hline
       \multicolumn{3}{c}{Content of SED files\tablefootmark{a}: \texttt{sed/??\textbackslash??\_sed\_\#\#\#\#\#.csv}\tablefootmark{b}}\\

    \texttt{nu\_Fnu}	& observed flux density & $erg\cdot s^{-1} \cdot cm^{-2}$\\
    \texttt{nu\_Fnu\_error} & flux density uncertainty & $erg\cdot s^{-1} \cdot cm^{-2}$\\
    \texttt{wavelength} & effective wavelength & \AA \\
    \texttt{flux} & observed flux & Jy \\ 
    \texttt{freq} & frequency & Hz\\ 
    \texttt{bib} & NASA/ADS \texttt{bibcode} for the original paper linked to the reported quantity & -\\ 
    \texttt{filter} & SVO filter name & -\\
    \hline
    \multicolumn{3}{c}{\underline{SED summary table} (\texttt{sed/sed\_summary.csv})} \\
    \texttt{uv}  & Number of SED datapoints in the range $10^{-3}- 0.4$\AA & -  \\
    \texttt{optical\_1} & Number of SED datapoints in the $\lambda$ range $0.4 - 0.6$\AA & - \\
    \texttt{optical\_2} & Number of SED datapoints in the $\lambda$ range $0.6- 0.8$\AA& - \\
    \texttt{optical\_3} & Number of SED datapoints in the $\lambda$ range $0.8- 1$\AA& - \\
    \texttt{nir\_1} & Number of SED datapoints in the $\lambda$ range $1.0-2$\AA & -  \\
    \texttt{nir\_2} & Number of SED datapoints in the $\lambda$ range $2-3$\AA & - \\
    \texttt{nir\_3} & Number of SED datapoints in the $\lambda$ range $3-4$\AA& -  \\
    \texttt{mir\_1} & Number of SED datapoints in the $\lambda$ range $4-5$\AA & - \\
    \texttt{mir\_2} & Number of SED datapoints in the $\lambda$ range $5-6$\AA & - \\
    \texttt{mir\_3a} & Number of SED datapoints in the $\lambda$ range $6-8$\AA & - \\
    \texttt{mir\_3b} & Number of SED datapoints in the $\lambda$ range $8-10$\AA & - \\
    \texttt{mir\_4} & Number of SED datapoints in the $\lambda$ range $10- 20$\AA & - \\
    \texttt{mir\_5} & Number of SED datapoints in the $\lambda$ range $20-30$\AA & - \\
    \texttt{fir\_1} & Number of SED datapoints in the $\lambda$ range $30-60$\AA & - \\
    \texttt{fir\_2} & Number of SED datapoints in the $\lambda$ range $60-100$\AA & - \\
    \texttt{fir\_3} & Number of SED datapoints in the $\lambda$ range $100-10^3$\AA & - \\
\hline
\end{tabular}
\end{table*}

\begin{table*}[]
    \caption{Summary of data types related to YSOs' IR classification. 
    \tablefoottext{a}{IR classes follow Tab.~\ref{tab:yso_classes}. When no $\alpha_\mathrm{IR}$ could be estimated, a label \texttt{not classified} is attributed instead.}
    \tablefoottext{b}{Possible values for \texttt{IR\_range}: For example, \texttt{>2} indicates a generic IR-range with available photometry from 2$\mu$m, and \texttt{3-8} indicates that photometry in the range 3--8$\mu$m was used.'}
    \tablefoottext{c}{Classification methods are summarized in Sect.~\ref{sec:ir_emission}. Possible values: \texttt{standard} for the standard classification scheme \citep{Lada:87a,Andre:93a}; \texttt{Hernandez} for \citet{2007ApJ...662.1067H}; \texttt{MGM} for classification as in \citet{2012AJ....144..192M}; \texttt{TD} for classification schemes focused on Transition discs; \texttt{Robitaille} for evolution classes derived with \citet{Robitaille2006} models, \texttt{misc} for miscellaneous classification schemes.}
    \tablefoottext{d}{Set to \texttt{True} if further information about IR-emission is available in the literature (for example, IR-excesses) but have not included in the literature. }
    \label{tab:app:disc}}
    \centering

    \begin{tabular}{llcrr}
   \hline\hline
    key & description & units & count & \# sources\\
    \hline
    \multicolumn{5}{c}{\underline{Standard Classification Scheme} (\texttt{ir/stnadard\_classification\_nemesis.csv})}\\
    \texttt{alpha\_2\_24} & $\alpha_\mathrm{IR}$ estimated in Sect.~\ref{sec:alphaindex}. & - & 25\,799 & 25\,799 \\
 
     \texttt{class\_2\_24} & Standard infrared classes derived in Sect.~\ref{sec:alphaindex}\tablefootmark{a} & - & 27\,879 & 27\,879 \\ 
     \texttt{lambda\_min\_2\_24} & Minimum wavelength available for $\alpha_\mathrm{IR}$ estimation & $\mu$m & 25\,799 & 25\,799 \\     
        \texttt{lambda\_max\_2\_24} & Maximum wavelength available for $\alpha_\mathrm{IR}$ estimation & $\mu$m & 25\,799 & 25\,799 \\
        
     \multicolumn{5}{c}{\underline{Literature values} (\texttt{disk\_literature.csv})}\\
\texttt{alpha} &  $\alpha_\mathrm{IR}$ compiled from the literature & - &  48\,657 & 5\,379 \\
\texttt{class} &  IR class compiled from the literature & - & 16\,716 & 8\,097\\
 \texttt{IR\_range} & wavelength range used for classification  & - & 62\,986 &  9\,438\\
\texttt{class\_method} & IR classification method\tablefootmark{c} & - & 62\,986 &  9\,438  \\
\texttt{disc} & disc description keyword & - & 8\,054 & 4\,096  \\
\texttt{IR\_ext} & Furhter infrared information available\tablefootmark{d} & - & 1\,159 & 1\,137 \\\hline
\end{tabular}
\end{table*}

\begin{table*}[]
    \caption{Summary of data types related to Lithium. \label{tab:Lithium}}
    \centering
    \begin{tabular}{lp{0.4\textwidth}crr}
    \hline\hline
    key & description & units & count & \# sources\\
    \hline
      \multicolumn{5}{c}{\underline{Lithium} (\texttt{lithium/li.csv})}\\
    
  \texttt{EW\_Li} & Equivalent Width at $\lambda$6708 \AA  & \AA & 8\,906 & 6\,337 \\
 \texttt{A\_Li} & Lithium abundance & dex & 2\,324 & 1\,997 \\\hline
\end{tabular}
\end{table*}

\begin{table*}[]
    \caption{Summary of data types related to gravity (see Sect.~\ref{sec:gravity} and Fig.~\ref{fig:logg}). \label{app:tab:gravity}}
    \centering
    \begin{tabular}{lp{0.4\textwidth}crr}
    \hline\hline
        key & description & units & count & \# sources\\
    \hline
    \multicolumn{5}{c}{\underline{$\log g$} (\texttt{gravity/logg\_DB.csv})}\\
    
  \texttt{logg} & $\log{g}$ derived from spectra fit& dex & 57\,649 & 12\,313
 \\
 \\
      \multicolumn{5}{c}{\underline{line measurements} (\texttt{gravity/ew\_gravity.csv})}\\
      \multicolumn{5}{l}{\underline{Na I doublet in absorption}} \\
  \texttt{NaI\_8200} & Equivalent width at $\lambda$8200 \AA  (unresolved doublet) & \AA & 1\,451 & 1\,362 \\
  \texttt{NaI\_8195} & Equivalent width at $\lambda$8195 \AA & \AA & 102 & 102 \\
      \multicolumn{4}{l}{\underline{K I doublet in absorption}}\\
  \texttt{EW\_KI\_7665} & Equivalent width at $\lambda$7665 \AA &  \AA & 52 & 52 \\
  \texttt{EW\_KI\_7699} & Equivalent width at $\lambda$7699 \AA &  \AA & 39 & 39 \\
    \texttt{EW\_KI\_7665\_7699} & Equivalent width at $\lambda\lambda$7665-7699 \AA &  \AA & 6 & 6 \\
      \multicolumn{4}{l}{\underline{CaH3 }}\\
  \texttt{EW\_CaH3\_6965} & Equivalent width at $\lambda$6965 \AA & \AA & 6 & 6 \\
       \multicolumn{4}{l}{\underline{TiO}}\\
  \texttt{EW\_TiO\_8442} & Equivalent width at $\lambda$8442 \AA & \AA & 218 & 218 \\
       \multicolumn{4}{l}{\underline{TiO5}}\\
  \texttt{EW\_TiO\_7130} & Equivalent width at $\lambda$7130 \AA & \AA & 6 & 6 \\  \hline
\end{tabular}
\end{table*}

\begin{table*}[]
    \caption{Summary of data types related to observables of mass inflow/outflow. \tablefoottext{a}{Qualitative flags describing the H$\alpha$ line profile as reported by \citet{2014AandA...570A..30P}.} \label{tab:acc}}
    \centering
    \begin{tabular}{lp{0.4\textwidth}crr}
    \hline\hline
        key & description & units & count & \# sources\\
    \hline
\multicolumn{5}{c}{\underline{H$\alpha$} (\texttt{accretion/Ha\_DB.csv})} \\
    \texttt{EW\_Ha} & EW of $H\alpha$ at $\lambda6563$ \AA & \AA & 32\,760 & 20\,667 \\
\texttt{FW$_{10\%}$(H$\alpha$)} & Full-width at 10$\%$ of the H$\alpha$ emission profile peak & km/s & 1\,181 & 945\\
    \texttt{EW\_Ha\_PHOT} & EW of $H\alpha$ from photometry & \AA & 1\,514 & 1\,239\\
\texttt{Ha\_line\_flag} & Flag for $H\alpha$ profile classification\tablefootmark{a} & - & 1\,731 & 1\,731 \\

\\
\multicolumn{5}{c}{\underline{other lines}(\texttt{accretion/EW\_accretion\_other\_DB.csv})} \\
\multicolumn{4}{l}{\underline{Other Hydrogen Lines}}\\
\texttt{EW\_Hb\_4861} & EW of H$\beta$  at $\lambda4861$ \AA & \AA & 2357 & 2319 \\
 \texttt{EW\_Hg\_4340} & EW of H$\gamma$ at $\lambda4340$ \AA & \AA & 57 & 48 \\
  \texttt{EW\_H11\_3771} & EW of H11 at $\lambda3771$ \AA & \AA & 8 & 8 \\
\texttt{EW\_Pab\_12822} & EW of Paschen $\beta$ at $\lambda12822$ \AA  & \AA & 15 & 15 \\
\texttt{EW\_Pag} &EW of Pa$\gamma$ ($\lambda1090$ \AA) & \AA & 7 & 7\\
 \texttt{EW\_Brg\_21661} & EW of Brackett $\gamma$ at $\lambda21661$ \AA & \AA & 29 & 29 \\
  \texttt{EW\_Br11\_16810} & EW of Brackett 11 at $\lambda16810$ \AA & \AA & 158 & 158 \\
\multicolumn{4}{l}{\underline{ Helium Lines}}\\
 
 \texttt{EW\_HeI\_6678} & EW of He I line at $\lambda$6678 \AA  & \AA & 249 & 231 \\
 \texttt{EW\_HeI\_5876} & EW of He I line at $\lambda$ 5876 \AA  & \AA & 63 & 57 \\
 \texttt{EW\_HeI\_10830} & EW of He 
 I line at $\lambda$ 10830 \AA  & \AA & 58 & 54 \\
\multicolumn{4}{l}{\underline{Calcium Lines}}\\

\texttt{EW\_CaII\_8498} & EW of Ca II triplet at $\lambda$8498 \AA  & \AA & 130 & 122 \\
\texttt{EW\_CaII\_8542} & EW of Ca II triplet at $\lambda$8542 \AA  & \AA & 1142 & 1134 \\
\texttt{EW\_CaII\_8662} & EW of Ca II triplet at $\lambda$8662 \AA)  & \AA & 135 & 129 \\
\texttt{EW\_CaII\_H\_3934} & EW of Ca II H$\&$K doublet at $\lambda$3964 \AA  & \AA & 49 & 41 \\
\texttt{EW\_CaII\_K\_3969} & EW of Ca II H$\&$K doublet at $\lambda$3969 \AA & \AA & 53 & 45 \\

\multicolumn{4}{l}{\underline{Forbidden Lines}}\\
 \texttt{EW\_\_OI\_\_5577} & EW of [O I] at $\lambda$5577 \AA   & \AA & 72 & 69 \\
  \texttt{EW\_\_OI\_\_6300} & EW of [O I] at $\lambda$6300 \AA   & \AA & 189 & 173 \\
   \texttt{EW\_\_OI\_\_6364} & EW of [O I] at $\lambda$6364 \AA   & \AA & 75 & 69 \\
   \texttt{EW\_\_SII\_\_6716} & EW of [S II] at $\lambda$6717\AA   & \AA & 162 & 157\\
 \texttt{EW\_\_SII\_\_6731} & EW of [S II] at $\lambda$6731\AA   & \AA & 169 & 160\\
 \texttt{EW\_\_FeII\_\_16440} & EW of [Fe II] at $\lambda$16440\AA   & \AA & 6 & 6\\
 \texttt{EW\_\_NII\_\_6548} & EW of [N II] at $\lambda$ 6548\AA   & \AA & 18 & 18 \\
  \texttt{EW\_\_NII\_\_6581} & EW of [N II] at $\lambda$ 6581\AA   & \AA & 16 & 16 \\
   \texttt{EW\_\_NII\_\_6583} & EW of [N II] at $\lambda$ 6583\AA   & \AA & 128 & 128 \\
\multicolumn{4}{l}{\underline{Other emission lines}}\\
 \texttt{EW\_OI\_7773} & EW of the OI at $\lambda$7773 \AA   & \AA & 49 & 49 \\
  \texttt{EW\_OI\_8446} & EW of the OI line at $\lambda$8446 \AA   & \AA & 54 & 54 \\
  \texttt{EW\_NaI\_5893} & EW of the NaI doublet at $\lambda$5893 \AA   & \AA & 8 & 8 \\
    \texttt{EW\_H2\_1\_0\_S(1)} & EW of the H$_2$ 1-0 S(1) emission at $\lambda$2.1218$\mu$m   & \AA & 13 & 13 \\
           \texttt{EW\_CO\_2\_0} & EW of the CO 2-0 1st overtone at $\lambda2.293\mu$m   & \AA &  24 & 24\\  %
\\
\multicolumn{5}{c}{\underline{Veiling} (\texttt{accretion/veiling\_DB.csv})} \\
\texttt{veiling\_r\_opt} & optical-veiling at 6390-6710\AA & - & 779 & 779 \\
\texttt{veiling\_r\_h} & ir-veiling at the H-band ($\sim1.5\mu$m) & - & 32\,132 & 9\,031 \\
\texttt{veiling\_r\_k} & ir-veiling at the K-band range ($\sim2.2\mu$m)& - & 3 & 3 \\
\texttt{veiling\_r\_r7465} & veiling around 7465\AA& - & 1\,444 & 361 \\
\end{tabular}
\end{table*}

\begin{table*}[]
    \caption{Summary of data types likely related to Radial Velocity.
    \label{tab:RV}
    }
    \centering
    \begin{tabular}{lp{0.4\textwidth}crr}
    \hline\hline
 key & description & units & count & \# sources\\
    \hline
     \texttt{RV} & RV & km/s & 60\,294 & 13\,721 \\ 
    \hline        
    \end{tabular}
\end{table*}

\begin{table*}[]
    \caption{Summary of data types likely related to stellar variability. \label{tab:var}}
    \centering
    \begin{tabular}{lp{0.4\textwidth}crr}
    \hline\hline
 key & description & units & count & \# sources\\
    \hline
    \multicolumn{5}{c}{\underline{Variability} (\texttt{variability/amplitude.csv})}\\

\texttt{AG\_proxy} & Gaia DR3 $A_G$-proxy variability index & - & 24\,632 & 24\,632\\\hline
    \end{tabular}
\end{table*}

\begin{table*}[]
    \caption{Summary of data types likely related to stellar rotation. \label{tab:Rot_data_types}}
    \centering
    \begin{tabular}{lp{0.4\textwidth}ccc}
    \hline\hline
 key & description & units & count & \# sources\\
 \hline
\multicolumn{5}{c}{\underline{Rotation} \texttt{(rotation/rotation.csv)}}\\

 \texttt{Per} & Variability periods & day & 21\,932 & 10\,147 \\
 \texttt{Per\_band} & Photometric band used for period measurement &- &- & -\\
     \texttt{vsini} & $v\sin{i}$ & km/s & 20\,820  & 10\,651 \\ 
    \hline        
    \end{tabular}
\end{table*}

\begin{table*}[]
    \caption{Summary of data types related to X-ray surveys (see Sect.~\ref{sec:xray}).
    \tablefoottext{a}{\texttt{HR\#\_band} possible values:\texttt{0.5-2.0/2.0-8.0keV}, 
\texttt{(1.0-2.0)-(0.5-1.0)/(0.5-2.0)keV}, 
\texttt{(2.0–4.5 keV)-(0.3–2.0 keV)/(2.0–4.5 keV)+(0.3–2.0 keV)}, 
\texttt{0.5-1.7/1.7-2.8keV}, 
\texttt{(1.0-7.3)-(0.5-1.0)/(0.5-7.3)keV}, 
\texttt{(4.5–7.5 keV)-(2.0–4.5 keV)/(4.5–7.5 keV)+(2.0–4.5 keV)},  
\texttt{1.7-2.8/2.8-8.0keV}
\tablefoottext{b}{\texttt{survey} possible values: \texttt{COUP},
\texttt{Chandra}, 
\texttt{XMM-Newton}, 
\texttt{Einstein}, 
\texttt{ROSAT}}
}\label{app:tab:xray}}
    \centering
    \begin{tabular}{lp{0.4\textwidth}ccc}
    \hline\hline
    key & description & unit & count & \# sources \\
    \hline
    \multicolumn{5}{c}{\texttt{xray/Xray\_DB.csv}} \\
    
 \texttt{Fx} & X-ray flux & erg/cm$^2$/s & 2\,371 & 2\,333\\
 \texttt{logLx} & Total X-ray luminosity & erg/s & 6\,121 & 3\,051 \\
 \texttt{\#\_band} & Energy band observed. \# paired with \texttt{Fx} or \texttt{logLx}. & keV & & \\
  \texttt{logNH} & Hydrogen column density  & $10^{21}cm^{-2}$ & 4\,324 & 1\,952 \\
\texttt{HR\#} & Hardness ratio \#, \# in (1, 2, or 3) &  - & 1\,364 & 1\,364\\
\texttt{HR\#\_band} & HR\# definition\tablefootmark{a}  & - &- & -\\
\texttt{survey} & Facility used for observations\tablefootmark{b} &  - & - & - \\
\hline
\end{tabular}
\end{table*}

\begin{table*}[]
    \caption{Summary of data types related to source multiplicity evaluation (Sect.~\ref{sec:multiplicity}).
    \tablefoottext{a}{Multiplicity labels follow discussion in Sect.~\ref{sec:multiplicity}: \texttt{S}: Binary identified from spectroscopy; \texttt{E}: Eclipsing binary; \texttt{A}: Astrometric binary; \texttt{B}: Generic binary; \texttt{V}: Visual pair; \texttt{U}: Unresolved pair; \texttt{Bl}: Blended source; \texttt{R}: RUWE unresolved candidate ; \texttt{?}: Candidate.}\label{app:bin}}
    \centering
    \begin{tabular}{lp{0.4\textwidth}crr}
    \hline\hline
    key & description & unit & count & \# sources \\
    \hline
    \multicolumn{5}{c}{\texttt{multiplicity/multiplicity\_label.csv}} \\
\texttt{bin\_type} & Multiplicity flag as in Sect.~\ref{sec:multiplicity}\tablefootmark{a}  & - & 18\,929 & 12\,155\\
\texttt{bin\_ext\_type} & Multiplicity type in source paper  & - & 18\,659 & 12\,102\\
\hline
\end{tabular}
\end{table*}

\section{Extragalactic and Giants contamination assessment}\label{app:galaxy}

To investigate extragalactic and giant contamination in our catalogue, we used \texttt{Scikit-learn} \citep{scikit-learn} to design two binary Random Forest classifiers \citep{2001MachL..45....5B} trained to disaggregate these types of sources based on a set of attributes and a labelled sample. Similar classifiers specialised in the separation of stellar and extragalactic sources have been extensively implemented in the literature \citep[e.g.,][]{Vasconcellos2011AJ....141..189V,LoganFotopoulou2020AA...633A.154L,Cook2024MNRAS.535.2129C,Lourens2024AA...690A.224L}. However, these previous studies do not explicitly include YSOs in their training sample. Hence, here, we build our own classifier tailored for the OSFC field.

\subsection{Data}

We built a reference photometric catalogue for the OSFC field 
based on large photometric surveys summarised in Sect.~\ref{sec:largePhotSurveys}, with a base list of 16\,177\,367 sources with data from Gaia DR3, PanSTARRS1-DR2 and CatWISE. Near-infrared data were excluded because there were no surveys covering the entire OSFC field with comparable depth to the other surveys. We refer to the RF classifier trained to identify extragalactic sources against stellar sources as ``Extragalactic classifier'', and the one trained to identify giant stars against stellar and extragalactic sources, as ``Giant classifier''.

\subsection{Labelled dataset}\label{app:RFlabels}

\subsubsection{Extragalactic sources}
\label{sec:GalLit}

We collated a sample of previously identified extragalactic sources, including a mixture of resolved and unresolved Galaxies, AGNs and QSOs. This accounts for 32\,850 sources in the OSFC field. To maximise purity, we kept only extragalactic sources with a match within $0.5''-1''$ to the photometric reference catalogue. The list of extragalactic sources was composed of sources from: 

\begin{itemize}
    \item{HyperLEDA database \citep{HyperLEDA}:} 2\,638 sources; 
    \item{Million Quasars catalogue v. 8 \citep{Flesch2023OJAp....6E..49F}:} 4\,641 sources;
    \item{Galaxy Zoo \citep[][]{Walmsley2023MNRAS.526.4768W}:} 6\,124 sources;
    \item{MANGROVE Catalogue \citep[][]{Biteau2021ApJS..256...15B}:} 1\,396 sources;
    \item{Gaia DR3 high-purity Galaxy sample:} 7\,718 sources;
    \item{Gaia DR3 high-purity QSO sample:} 15\,629 sources.
    \end{itemize}

\noindent Extragalactic sources from Gaia DR3 were identified by Gaia's ML classification and redshift estimation \citep{Delchambre2023AA...674A..31D}, brightness profile analysis \citep{Ducourant2023AA...674A..11D}, or variability behaviour \citep{Carnerero2023AA...674A..24C,Rimoldini2023AA...674A..14R}. High-purity samples were devised based on selection procedures outlined by \citet{GDR3_Extragalactic_2023AA...674A..41G}. 

\subsubsection{Stellar sources}\label{app:RF_stellar_labels}

The following samples of stellar sources were included in our labelled sample:
\begin{itemize}
    \item \emph{Certainly young}: 3\,698 sources from the NEMESIS YSO catalogue which have spectroscopic confirmation of youth based on significant Lithium absorption in their spectra (Sect.~\ref{sec:lithium});
    \item \emph{At Orion's distance}: 59\,146 Gaia DR3 sources with parallax consistent with a location around the location of the OSFC (300-500 pc) and parallax uncertainty better than 0.05 mas;
    \item \emph{Giants}:  13\,336 sources with $-0.5\leq\log{g}\leq2.5$ in either RAVE, LAMOST LRS or MRS (Sect.~\ref{sec:spec-surveys}), and  $\log{g}$ precision better than 0.25 dex;
    \item \emph{Other Stellar Sources:} 72\,336 sources with $\log{g}>3.5$ in either RAVE, LAMOST LRS or MRS (Sect.~\ref{sec:spec-surveys}), and  $\log{g}$ precision better than 0.25 dex.
\end{itemize}

\subsubsection{Final sample}

The labelled data were further processed to remove any overlapping labels. For example, sources labelled both \emph{Certainly Young} and \emph{Giant} were removed. 
Further pre-cleaning of the data included removal of sources with very large uncertainties.

\paragraph{Extragalactic classifier sample:}  All sources listed in Sect. \ref{sec:GalLit} were initially labelled as ``extragalactic'' (positive label), and all sources in Sect. \ref{app:RF_stellar_labels} were labelled as ``stellar'' (negative label). Validation of labelled samples was carried out by comparing the locus occupied by the different samples in colour-colour diagrams. During this procedure, we verified a significant tail of ``Gaia DR3 high-purity QSOs'' sources towards the location of stellar sources, which did not exist in the distribution of other samples with large numbers of QSOs (e.g., Million Quasars Catalogue). Furthermore, later classifier implementation tests revealed that the exclusion of ``Gaia DR3 high-purity QSO'' labelled greatly improved the classifier recall. We therefore excluded labels from the ``Gaia DR3 high-purity QSO'' sample from our final implementation. Further evidence that this sample suffers from contamination by YSOs is discussed in Sect.~\ref{sec:galaxy}. A final parent sample with 23\,217 and 94\,248 sources labelled as extragalactic and stellar was available for the Extragalactic classifier.

\paragraph{Giant classifier sample:}  Sources in the Giants sample in Sect.~\ref{app:RF_stellar_labels} were labelled as such (positive label), and extragalactic sources in Sect.~\ref{sec:GalLit} or in the ``certainly young'' and ``other stellar'' samples were labelled as ``other'' (negative label).  We deliberately avoid the sample at Orion's distance as this may contain Giants with unavailable $\log{g}$ measurements.  A final parent sample with 12\,988  and 133\,966 sources labelled as giant and others was available for the Giant classifier.

The final labelled datasets were split into 30\% training, 40\% validation and 30\% testing sets. We ensured samples contained a representative fraction of the different source types (e.g., resolved and unresolved galaxies) by stratifying them based on origin catalogues (Sect. \ref{sec:GalLit}), maintaining similar source proportions from each catalogue. Similarly, each subtype of stellar sources in Sect. \ref{app:RF_stellar_labels} was also adequately represented.

\subsection{Features}

Implementation tests were carried out by training classifiers in six subsets of features, combining available colours from Gaia DR3, PanSTARRS1-DR2, and CatWISE. We included both pure-colour subsets and versions with Gaia parallaxes and uncertainties. A summary of features used in each of our best-performing models is presented in Tab.~\ref{tab:features}.

\begin{table*}[htb]
    \centering
    \caption{Summary of features and average-precision score for the best-performing Random Forest Classifiers used for Extragalactic and Giant contamination evaluation. \emph{Model}: Internal name used to label the classifier; \emph{Classifier}: Aim of the classifier; \emph{\# features}: number of features used; \emph{Features}: set of features used; \emph{sources:} number of sources labelled by this classifier.\tablefoottext{a}{\texttt{colours()} refer to every valid photometric colour combination for the set within \texttt{()}.}}
    \label{tab:features}    
    \begin{tabular}{p{0.11\textwidth}ccp{0.5\textwidth}c}
    \hline\hline
       Model  & Classifier & \# features & features & sources\\
\hline
         \hline
        GaiaPS1WISE \_pure\_colour & Extragalactic & 40  & colours($W1_\mathrm{c}$, $W2_\mathrm{c}$,  {$G$}, {$G_{BP}$}, {$G_{RP}$}, $g_\mathrm{PS1}$, $r_\mathrm{PS1}$, $i_\mathrm{PS1}$, $z_\mathrm{PS1}$,  $y_\mathrm{PS1}$)\tablefootmark{a} & 6\,520\,467\\
        \hline
         Gaia & Giant &  5 & $G-G_{BP}$, $G_{BP}-G_{RP}$, $G- G_{RP}$,  \texttt{parallax}, \texttt{parallax\_error} & 8\,385\,766\\
         \hline        
    \end{tabular}
\end{table*}

\subsection{Training}

Class imbalance was intentionally maintained for both classifiers, as we expect giants and extragalactic sources to be a minority of sources in the OSFC field. However, this choice led to true negative labels dominating precision and accuracy. To reduce its impact on the smaller classes (true positive labels), we prioritised the use of scores that account for the trade-off between precision and recall for model comparison and performance evaluation. Additionally, as a general rule, models with validation sample true positive recovery rate under 75\% were discarded. Grid search with 5-fold cross-validation was performed to optimise hyperparameters, using average precision as the reference score. Each classifier was then trained with the optimal parameters and exported.

\subsection{Validation}

Validation samples were used for the comparison of the performance of each classifier trained, and a series of scores (namely, accuracy, precision, recall, F1-score and F$\beta(\beta=2$) score) was used to rank their performance. For the extragalactic classifier, we found that sets of attributes covering colours from optical to mid-infrared (\texttt{GaiaPS1WISE\_pure\_colour}) performed better, qualitatively consistent with results in previously published classifiers. For the giant classifier, the models using Gaia parallax performed better, with the \texttt{Gaia} model (attributes: Gaia colours plus parallax and its uncertainties) being only marginally outperformed by the \texttt{GaiaWISE} model (same Gaia attributes with the addition of CatWISE-based colours). Nevertheless, the requirement for both Gaia and CatWISE data for the latter makes it applicable to a smaller number of sources. We therefore opt to keep the results for the pure-Gaia model (Tab.~\ref{tab:features}).

Validation samples were also used to study the precision-recall curves of the best-performing classifiers. At this stage, we used the F1-score for determining the best classification thresholds while ensuring a good trade-off between precision and recall. Best classification thresholds were found at 0.41 for the extragalactic classifier, and at 0.39 for the giant classifier.

\subsection{Classification}

\begin{figure}
    \centering
    \includegraphics[width=0.49\linewidth]{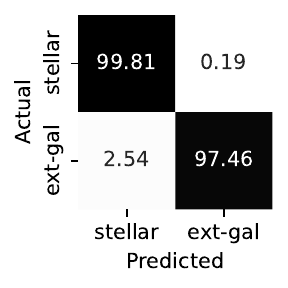}
    \includegraphics[width=0.49\linewidth]{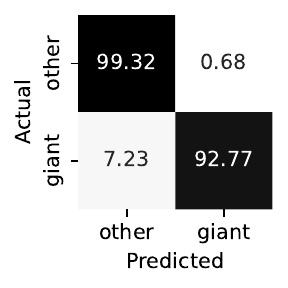}  \\
    \caption{Confusion matrices for the Extragalactic (Left) and Giant (Right) Random Forest classifiers normalised over true labels.}
    \label{fig:confusion_matrix_GalGiant}
\end{figure}

We used the classification thresholds from the previous section for classification. Classification reports for the testing set are presented in Tab.~\ref{tab:GalaxyGiant}, and confusion matrices in Fig.~\ref{fig:confusion_matrix_GalGiant}.

\begin{table}[htb]
\caption{Classification performance report of the Random Forest Extragalactic (top) and Giant (bottom) classifiers.}\label{tab:GalaxyGiant}
\centering
\begin{tabular}{rccc}
    \hline\hline
 label & precision & recall & F1-score \\
    \hline
    \multicolumn{4}{c}{Extragalactic classifier} \\
         stellar        & 1.00 &   1.00 &   1.00 \\
         extragalactic  & 0.99 &   0.97 &   0.98 \\
\hline
       \multicolumn{4}{c}{Giant Classifier}\\
          Other     & 0.99 & 0.99 & 0.99\\
 Giants   &  0.94 & 0.93 & 0.93\\
    \hline
   \end{tabular}
\end{table}

\begin{figure}
    \centering
    \includegraphics[width=0.9\linewidth]{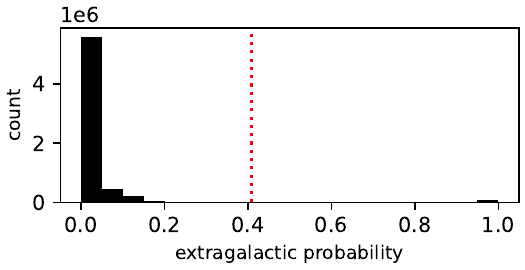}
    \includegraphics[width=0.9\linewidth]{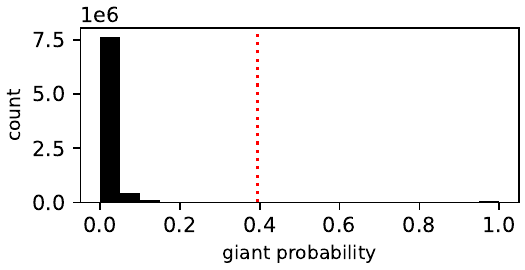} \\
    \caption{Distribution of probabilities of sources in the field of the OSFC being extragalactic (top) and giants (bottom). The red dotted line shows the threshold for classification.}
    \label{fig:PorbDist_Gal_Giant_RF}
\end{figure}

\paragraph{Extragalactic classifier:} We were able to apply the \texttt{GaiaPS1WISE\_pure\_colour} model to classify 6\,520\,467 sources, with 139\,668 classified as probable extragalactic sources. We note that among stellar sources misclassified as extragalactic (0.19 false negative rate), $\sim35\%$ were spectroscopically confirmed YSOs, although these correspond to only $\sim3\%$ of the YSOs in the testing sample.

\paragraph{Giant Classifier:}  We were able to apply the \texttt{Gaia} model to derive the probabilities of being a giant for 8\,385\,766 sources in the field of the OSFC, with 141\,808 likely giants identified. Among sources misclassified as giants in the testing sample, 97\% were labelled ``other stellar sources`` based on their $\log{g}$ measurements from spectroscopic surveys. The remaining 3\% were real YSOs with spectroscopic constraints, these correspond to $\sim 0.6 \%$ of the YSO sources in the testing sample.

The development of these classifiers included 53\,154 labelled sources without enough data for classification. Both samples of classified and labelled but not classified sources were matched to the OSFC YSO catalogue and further discussed in Sect.~\ref{sec:contamination}.

\end{appendix}
\end{document}